\documentclass{article}

\usepackage{arxiv}
\usepackage[utf8]{inputenc} 
\usepackage[T1]{fontenc}    
\usepackage[colorlinks,citecolor=blue,urlcolor=red]{hyperref}
\usepackage{url}            
\usepackage{booktabs}       
\usepackage{amsfonts}       
\usepackage{nicefrac}       
\usepackage{microtype}      
\usepackage{graphicx}
\usepackage{natbib}
\usepackage{doi}
\usepackage{amsmath}
\usepackage{color}

\newcommand{\magenta}{\color{magenta}}
\usepackage{comment}

\def\mybox#1{{\magenta \vskip1mm \begin{center}
 \hspace{.0\textwidth}\vbox{\hrule\hbox{\vrule\kern6pt
  \parbox{.8\textwidth}{\kern6pt#1\vskip6pt}\kern6pt\vrule}\hrule}
  \end{center} \vskip-5mm}}

\title{Comparative Effectiveness Research with Average Hazard for Censored Time-to-Event Outcomes: A Numerical Study}

\date{} 					

\author{ 
    {Hong~Xiong} \\ 
	Department of Biostatistics \\
    Harvard Chan School of Public Health \\ 
    Boston, Massachusetts 02115, USA 
	\And
    {Jean~Connors} \\ 
	Department of Hematologic Oncology\\
	Dana Farber Cancer Institute\\
	Boston, Massachusetts 02215, USA \\
	\And
    {Deb~Schrag} \\ 
	Department of Medicine\\
	Memorial Sloan Kettering Cancer Center\\
	New York, New York 10065, USA \\
	\And
    {Hajime~Uno}\thanks{Email: huno@ds.dfci.harvard.edu} \\
	Department of Data Science\\
	Dana Farber Cancer Institute\\
	Boston, Massachusetts 02215, USA \\
}

\renewcommand{\headeright}{Xiong and Uno}
\renewcommand{\undertitle}{}
\renewcommand{\shorttitle}{CER using Average Hazard}

\hypersetup{
pdftitle={AHadjust1},
pdfsubject={stat.app},
pdfauthor={Xiong and Uno},
pdfkeywords={AH, CER},
}

\begin{document}
\maketitle

\baselineskip=12pt

\begin{abstract}
The average hazard (AH), recently introduced by Uno and Horiguchi, represents a novel summary metric of event time distributions, conceptualized as the general censoring-free average person-time incidence rate on a given time window, $[0,\tau].$ This metric is calculated as the ratio of the cumulative incidence probability at $\tau$ to the restricted mean survival time at $\tau$ and can be estimated through non-parametric methods. The AH's difference and ratio present viable alternatives to the traditional Cox's hazard ratio for quantifying the treatment effect on time-to-event outcomes in comparative clinical studies. While the methodology for evaluating the difference and ratio of AH in randomized clinical trials has been previously proposed, the application of the AH-based approach in general comparative effectiveness research (CER), where interventions are not randomly allocated, remains underdiscussed. 
This paper aims to introduce several approaches for applying the AH in general CER, thereby extending its utility beyond randomized trial settings to observational studies where treatment assignment is non-random.
\end{abstract}


\baselineskip=12pt

\section{Introduction}

For decades, the hazard ratio, derived from Cox's proportional hazards (PH) model \citep{Cox:1972un}, has been the predominant metric for quantifying treatment effects in time-to-event outcomes in clinical research. 
However, despite its statistical elegance and efficiency, the traditional Cox's hazard ratio (HR) approach has significant limitations that hinder its ability to meet crucial and practical objectives in estimating treatment effect magnitudes \citep{Hernan2010-fc,Royston:2011ci,Royston:2013ho,Uno:2014ii,Uno:2015ip}. 
Specifically, the CONsolidated Standards of Reporting Trials (CONSORT) 2010 guideline \citep{Cobos-Carbo2011-wz} and the Annals of Internal Medicine's General Statistical Guidance (\citeauthor{annals}, 2024) emphasize the importance of presenting treatment effect magnitudes in both absolute and relative terms to aid interpretation. However, the inherently relative nature of Cox's HR often precludes its conversion into an absolute hazard difference (except in some special cases), thus failing to meet this important criterion for practical interpretation.
Additionally, the Cox model's reliance on the PH assumption poses another limitation, as this assumption frequently does not hold in various contexts, including certain cancer immunotherapy trials. When the PH assumption is violated, the resulting HR estimates derived from the traditional Cox's HR approach are influenced by study-specific censoring time distributions, making it difficult to generalize them to future patient populations \citep{Kalbfleisch1981-xo,Struthers:1986uy,Lin:1989vg,Xu:2000kja,Horiguchi2019-xc}.

Given the limitations of this approach, the exploration and adoption of various alternative approaches has been discussed. Among these alternative approaches, one promising class uses a summary measure of event time distribution, such as median, the cumulative incidence probability at a specific time point, and the restricted mean survival time (RMST) for each group and then constructs the between-group metrics, such as ratio and difference \citep{Uno:2014ii}. 
The approaches in this class do not impose any strong model assumptions on the relationship between the intervention groups, and thus provide robust estimates for the magnitude of the treatment effect. In addition, the summary measures of the event time distributions from two groups easily allow for the expression of the treatment effect in both absolute difference and relative terms. This will aid in enhancing interpretation of the treatment effect magnitude, meeting guideline recommendations (\citeauthor{Cobos-Carbo2011-wz}, 2011; \citeauthor{annals}, 2024). 

While the RMST-based approach in this class is gaining more attention 
\citep{Royston:2011ci,Royston:2013ho,Uno:2014ii,Uno:2015ip} 
and is beginning to be used in practice \citep{Guimaraes2020-zf,Connolly2022-sx, Hammad2022-ap, Sanchis2023-bl}, 
the average hazard with survival weight (AH) has emerged more recently as another promising approach in this class \citep{Uno2023-sm}. 
The AH provides a new perspective by expressing the average incidence of person-time events over a specified time window without the nuisance censoring time distribution affected in general. 
Specifically, the AH is defined as 
\begin{equation} 
\eta(\tau) 
= \frac{\int_{0}^{\tau} h(u)S(u)du}{\int_{0}^{\tau} S(u)du} 
= \frac{E\{I(T\le \tau)\}}{E\{ T \wedge \tau \}},
\label{AH}
\end{equation}
where $\tau$ is a truncation time point, $h(u)$ and $S(u)$ are the hazard function and survival function for the event time $T,$ respectively, 
$I(A)$ is an indicator function for the event $A,$
and $x \wedge y $ denotes $\min(x,y).$
The AH is closely related to the person-time incidence rate (IR) commonly defined as {\it `the number of the observed events divided by the total observation time'} \cite{rothman2008modern}, that is, 
\begin{equation} 
IR 
= \frac
  {\sum_{i=1}^{n}I(T_i \le C_i)}
  {\sum_{i=1}^{n}( T_i \wedge C_i )},
\label{IR}
\end{equation}
where $n$ is the sample size of the analysis population, and $\{ T_i \}$ and $\{  C_i \} $ for $i=1 \ldots, n$ are the independent copies from the event time $T$ and censoring time $C,$ respectively. As $n$ goes to $\infty,$ this converges to \begin{equation} 
 \frac
  {E\{ I(T \le C) \} }
  {E\{  T \wedge C \} }.
\label{IRlimit}
\end{equation}
 Note that this quantity depends on the censoring time distribution, except in some special cases such that the event time $T$ follows an exponential distribution. On the other hand, (\ref{AH}) does not involve $C$ but a fixed time point $\tau.$ 
 Because the censoring time distribution is study-specific and nuisance for the inference about $T,$ we prefer the AH with a specified truncation time $\tau$ to (\ref{IRlimit}). 
 In contrast to the conventional person-time incidence rate, (\ref{IR}) or (\ref{IRlimit}), the AH can also be called {\it the general censoring-free incidence rate (gCFIR)} on a time window $[0,\tau].$  
 
For a comparison between Group 0 and 1, difference in AH (DAH) and ratio of AH (RAH) can be defined as 
\begin{equation} 
 \eta_1(\tau)-\eta_0(\tau) 
=  \frac{1-S_1(\tau)}{\int_{0}^{\tau} S_1(u)du}
 - \frac{1-S_0(\tau)}{\int_{0}^{\tau} S_0(u)du},
\label{DAH}
\end{equation}
and
\begin{equation} 
 \eta_1(\tau)/\eta_0(\tau) 
=  \left\{ \frac{1-S_1(\tau)}{1-S_0(\tau)} \right\} 
   \left\{ \frac{\int_{0}^{\tau} S_0(u)du}{\int_{0}^{\tau} S_1(u)du}\right\},
\label{RAH}
\end{equation}
respectively, where $S_k(t)$ is the survival function for the group $k.$
A non-parametric approach for the inference of DAH and RAH is to plug in the Kaplan-Meier (KM) estimates for $S_1(t)$ and $S_0(t)$ in equations (\ref{DAH}) and (\ref{RAH}). This approach has already been investigated under randomized trials settings by \cite{Uno2023-sm}. An R package (survAH) is also available for implementation of the KM-based inference. 

Currently, the application of the AH-based approach to general comparative effectiveness research (CER), where interventions are not allocated through randomization, has yet to be well investigated. In general within CER, the KM-based approach previously outlined for calculating DAH and RAH falls short in adequately adjusting for variations in case-mix between groups and can result in biased estimates of differences between groups.
However, it is evident from equations (\ref{DAH}) and (\ref{RAH}) that DAH and RAH are derived as functions of the survival functions of the respective groups. This suggests that the AH approach can be seamlessly integrated into CER by combining it with existing methods developed to derive {\it adjusted} survival functions \citep{chang1982corrected, makuch1982adjusted, cole2004adjusted, xie2005adjusted, austin2014use, wang2019bias, robins1992recovery, hubbard2000nonparametric}.
Yet, the effectiveness of this combined approach, particularly its numerical performance when applied to AH, remains underexplored.

This paper seeks to address the existing gap by merging established techniques for deriving adjusted survival functions with the AH-based approach in CER. 
Through extensive numerical studies, we evaluated the statistical performance of these integrated methods. 
We explored the practical applicability of these approaches for real-world CER studies, supplemented by an illustrative example using real data from an effectiveness study evaluating direct oral anticoagulants (DOAC) compared with low-molecular-weight heparin (LMWH) for the prevention of recurrent venous thrombosis (VTE) events in cancer patients who experienced an initial VTE event \citep{Schrag2023-ym}.

\section{Methods and Materials}

\subsection{Overview}

The objective of our numerical studies was to systematically evaluate the performance of various statistical methods for adjusting survival curves under different experimental settings when performing statistical inference regarding the causal treatment effects on survival time outcomes using AH as the summary measure of the survival time distribution. 

For each experimental setting, survival data were generated and analyzed by various confounding adjustment methods with details described in the next subsection. We employed several specialized R packages, including the \textit{survival}, \textit{survAH}, \textit{riskRegression}, \textit{pammtools}, \textit{Matching}, \textit{geepack}, and \textit{adjustedCurves} packages.

For each analytical method, the adjusted survival curves 
$\hat{S}_1(t)$ and $\hat{S}_0(t)$ were computed for the treatment (Group 1) and control (Group 0) groups, respectively. The adjusted AH's were calculated as
$\hat{\eta}_1(\tau) = \left\{1-\hat{S}_1(\tau)\right\}/{\int_{0}^{\tau} \hat{S}_1(u)du}$ and
$\hat{\eta}_0(\tau) = \left\{1-\hat{S}_0(\tau)\right\}/{\int_{0}^{\tau} \hat{S}_0(u)du},$ 
respectively.
The causal treatment effect parameters interest, DAH and RAH, were then computed by
$ \hat{\eta}_1(\tau) - \hat{\eta}_0(\tau)$
and
$ \hat{\eta}_1(\tau) / \hat{\eta}_0(\tau),$ respectively.
Corresponding 0.95 confidence intervals were also computed using bootstrap. 
We chose bootstrap for our numerical studies as not all confounding adjustment methods included had analytical variance available. Adopting bootstrap confidence intervals for all methods, regardless of the availability of analytic variance formula or software, is a standardized approach for constructing confidence intervals and allows for a fair performance comparison for the adjustment methods.

This process from data generation to analysis was repeated 2000 times. The final step of our simulation studies involved collating and summarizing the results across all iterations, evaluating each method through statistical performance metrics regarding inference of DAH, RAH, ${\eta}_1(\tau),$ and ${\eta}_1(\tau).$

\subsection{Confounding Adjustment Methods Outline}

The following six confounding adjusting methods were selected based on \cite{denz2023comparison} and included in our simulation study: 1) Direct Standardization, 2) Inverse Probability of Treatment Weighted Kaplan-Meier Estimates, 3) Inverse Probability of Treatment Weighted Survival via Cumulative Hazard, 4) Propensity Score Matching, 5) Empirical Likelihood Estimation, and 6) Augmented Inverse Probability of Treatment Weighting. 
Due to our focus on main approaches, we excluded methods using pseudo-values as they are not commonly used for confounding adjustment. 
Additionally, the standard Kaplan-Meier approach was included as a reference to demonstrate the effect of the absence of confounding adjustment. A brief introduction to each method is given below. 

\textbf{1) Direct Standardization}

The Direct Standardization (DS) method, or ``G-Formula'', estimates adjusted survival curves via the Cox regression model \citep{chang1982corrected, makuch1982adjusted,robins1986new}. 
This method fits the data to a Cox regression model, including the treatment group indicators and confounding factors. 
Based on the fitted model, the predicted survival curves are derived for all subjects in the dataset, assigning them to Group 1. The adjusted survival curve for Group 1 is then given by the average of these predicted survival probabilities. This is then repeated in the same manner to give the adjusted survival curve for Group 0. The validity of this approach relies on the adequacy of the specified model to predict the time-to-event outcome. 

\textbf{2) Inverse Probability of Treatment Weighted Kaplan-Meier Estimates} 

In the Inverse Probability of Treatment Weighted Kaplan-Meier Estimates (IPTW KM) approach, the Kaplan-Meier estimator is applied to samples that are weighted using the Inverse Probability of Treatment Weighting (IPTW) method to yield adjusted survival curves. The IPTW methodology is designed to address confounding by modeling the mechanism of treatment assignment \citep{Robins2000-so}. Typically, logistic regression is employed to calculate propensity scores for each individual across different levels of treatment \citep{Rosenbaum1983-dg}. Subsequently, individuals are assigned weights based on these estimated propensity scores. The Kaplan-Meier estimator is then used on these weighted samples to generate adjusted survival curves \citep{xie2005adjusted}. Assuming accurate modeling and estimation of weights, this method effectively mitigates confounding, leading to unbiased estimates.

\textbf{3) Inverse Probability of Treatment Weighted Survival via Cumulative Hazard} 

The Inverse Probability of Treatment Weighted Survival via Cumulative Hazard (IPTW CH) approach was introduced by \citet{cole2004adjusted}. It utilizes a stratified Cox regression model in conjunction with IPTW. In the Cox model, confounding factors are not included; instead, the model includes only the treatment group indicator as a stratification factor and utilizes IPTW for individual weighting. Statistical software, such as the {\it coxph} function in R and {\it PROC PHREG} in SAS, is used to calculate the adjusted cumulative hazard function for each treatment group. The adjusted survival function is subsequently obtained by exponentiating the negative of this adjusted cumulative hazard function.

\textbf{4) Propensity Score Matching}

Propensity Score Matching (Matching) represents an alternative, well-established method to control for confounding by employing the application of propensity scores \citep{Rosenbaum1983-dg}. This approach deviates from the use of propensity scores for individual weighting. Instead, it involves the creation of a new dataset, composed of individuals who exhibit similar estimated propensity scores. This methodology aims to achieve a balanced distribution of confounding factors across treatment groups. Subsequently, the Kaplan-Meier estimator is applied to these matched samples, enabling the derivation of adjusted survival curves. Assuming accurate specification of the propensity scores and the efficacy of the matching algorithm, the resultant survival curves are expected to be unbiased \citep{austin2014use}.

\textbf{5) Empirical Likelihood Estimation}

Empirical Likelihood Estimation (EL) represents a non-parametric approach grounded in likelihood estimation \citep{wang2019bias}. 
By maximizing a constrained likelihood function, the EL methodology aims to align moments of covariates between treatment groups, thus achieving similarity in distributional characteristics and removing bias. Notably, given the absence of a specified outcome model and treatment assignment model, the EL approach exhibits enhanced robustness to potential model misspecifications compared to methods necessitating a model, such as IPTW.

\textbf{6) Augmented Inverse Probability of Treatment Weighting}

Augmented Inverse Probability Weighting (AIPTW) approach marks a notable advancement in statistical methodologies for causal inference. Initially proposed by \citet{robins1992recovery}, it has been further refined by various researchers in different contexts \citep{hubbard2000nonparametric,zhang2012contrasting, ozenne2020estimation, bai2013doubly}. This method distinctively integrates both the outcome and treatment models, leveraging the strengths of each to enhance the estimation process. Specifically, AIPTW utilizes the outcome model to enhance the IPTW estimate based on the treatment model, with the goal of improving its efficiency. This approach ensures asymptotic unbiasedness, provided that at least one of the models --- either the outcome model or the treatment model --- is correctly specified. This grants AIPTW its notable doubly robust property. This feature is a significant strength of AIPTW, offering a robust layer of protection in the face of potential inaccuracies in model specification.

\textbf{7) Standard Kaplan-Meier method}

The Kaplan-Meier (KM) method offers a non-parametric approach to estimate the survival function from time-to-event data \citep{kaplan1958nonparametric}. 
In this approach, we estimate the survival functions for Group 0 and Group 1, separately. This unadjusted approach was included in the numerical studies as a reference.

\subsection{Data Description} 
\subsubsection{Data Generation Procedure for Covariates and Treatment Group Indicator} 

We followed the settings from \cite{denz2023comparison} to generate the simulation datasets. 
First, we generated the covariate vector   
${\bf X}=(X_1,\ldots,X_6)^\prime.$
Specifically,
$X_1 \sim \text{Bernoulli}(0.5),$ $X_2 \sim \text{Bernoulli}(0.3 + X_3 \cdot 0.1),$ $X_3 \sim \text{Bernoulli}(0.5),$ $X_4 \sim \mathcal{N}(0, 1),$ $X_5 \sim 0.3 + X_6 \cdot 0.1 + \mathcal{N}(0, 1),$ and $X_6 \sim \mathcal{N}(0, 1),$
where $\mathcal{N}(0,1)$ denotes the standard normal distribution.
We then generated the treatment group indicator $Z$ from a Bernoulli distribution with probability 
$p = \exp(a)/\left\{ 1+\exp(a)\right\},$ where  
$$ a = -1.2 + \log(3) \cdot X_2 + \log(1.5) \cdot X_3 + \log(1.5) \cdot X_5 + \log(2) \cdot X_6. $$

\subsubsection{Data Generation Procedure for Event Time} 

Let $U$ denote a random variable that follows a uniform distribution ranged from 0 to 1. For the event time $T,$ the following two scenarios were considered: 
\begin{equation*}
\begin{split}
\mbox{Scenario 1: }
T & = \left( \frac{-\log U}{\exp(\log(1.8) \cdot X_1 + \log(1.8) \cdot X_2 + \log(1.8) \cdot X_4 + \log(2.3) \cdot X_5^2 - 1 \cdot Z)} \right)^{0.5} \\ 
\mbox{Scenario 2: }
T & = \left( \frac{-\log U}{\exp(\log(1.8) \cdot X_1 + \log(1.8) \cdot X_2 + \log(1.8) \cdot X_4 + \log(2.3) \cdot X_5^2 - 1 \cdot Z)} \right).
\end{split}
\end{equation*}

\subsubsection{Data Generation Procedure for Censoring Time} 
The censoring time, $C,$ was generated independently of $T.$ Two censoring patterns were considered as follows:
\begin{equation*}
\begin{split}
\mbox{Censoring pattern A: }
C & = \min(C^*, 1.0), C^* \sim \mbox{Exp}(0.1) \\
\mbox{Censoring pattern B: }
C & = \min(C^*, 1.0), C^* \sim \mbox{Exp}(0.2), \\
\end{split}
\end{equation*}
where $\mbox{Exp}(\lambda)$ denotes the exponential distribution with the rate parameter 
$\lambda.$

\subsubsection{Derivation of Analysis Datasets}
For a given sample size $n,$ we generated $\{(T_i,C_i,{\bf X}_i,Z_i); i=1,\ldots,n\}$ and calculated the observable data $\{(Y_i,\Delta_i,{\bf X}_i,Z_i); i=1,\ldots,n\}, $
where $Y_i = \min(T_i,C_i),$ and $\Delta_i$ was 1 if $T_i \leq C_i$ and 0 otherwise. For each simulation configuration, we generated 2000 sets of $\{(Y_i,\Delta_i,{\bf X}_i,Z_i); i=1,\ldots,n\}$ to evaluate the statistical performance of the methods described in Section 2.2.

\subsection{Analyses}
\subsubsection{Cases}
Most of the methods introduced in Section 2.2 require users to specify either the outcome model, the treatment model, or both. Although it is preferred that the correct model is specified, this is not always the case in practice. 
Therefore, we considered the following five different cases regarding model specification by users.

\textbf{Case 1: Both models correct} \\
In Case 1, we assumed that both the outcome model and treatment model were correctly specified. Let $h(t | Z, {\bf X})$ be the hazard function for the event time $T,$ given $Z$ and $\bf X,$ and let $E[Z = 1 | {\bf X}]$ denote the expected value of getting the treatment for the subject with the covariate vector ${\bf X}.$ 
The outcome model and treatment model in Case 1 are then given by 
\begin{align*}
h(t | Z, {\bf X}) &= h_0(t) \exp(\beta_0 Z + \beta_1 X_1 + \beta_2 X_2 + \beta_4 X_4 + \beta_5 X_5), \\
E[Z = 1 | {\bf X}] &= g(\alpha_0 + \alpha_2 X_2 + \alpha_3 X_3 + \alpha_5 X_5 + \alpha_6 X_6), 
\end{align*}
respectively, where $h_0(t)$ is the baseline hazard function and $g(x) = {\exp(x)}/\left\{1+\exp(x)\right\}.$

\textbf{Case 2: Both models wrong by adding extra variables} \\
In Case 2, we assumed that users included extra variables in both outcome and treatment models, and thus neither the outcome model nor the treatment model was correctly specified. 
Specifically, the outcome model included $X_3$ and $X_6$, although these were not associated with the outcome. The treatment model included $X_1$ and $X_4$ that were not associated with $Z.$ 
\begin{align*}
h(t | Z, {\bf X}) &= h_0(t) \exp(\beta_0 {Z}+\beta_1 X_1 + \beta_2 X_2 + \beta_3 X_3 + \beta_4 X_4 + \beta_5 X_5 + \beta_6 X_6), \\
E[Z = 1 | {\bf X}] &= g(\alpha_0 + \alpha_1 X_1 + \alpha_2 X_2 + \alpha_3 X_3 + \alpha_4 X_4 + \alpha_5 X_5 + \alpha_6 X_6).
\end{align*}

\textbf{Case 3: Wrong outcome model and correct treatment model} \\
In Case 3, we assumed that users misspecified the outcome model with $X_2$ omitted but were able to specify the treatment model correctly. 

\begin{align*}
h(t | Z, {\bf X}) &= h_0(t) \exp(\beta_0 {Z} + \beta_1 X_1 + \beta_4 X_4 + \beta_5 X_5) \\
E[Z = 1 | {\bf X}] &= g(\alpha_0 + \alpha_2 X_2 + \alpha_3 X_3  + \alpha_5 X_5 + \alpha_6 X_6).
\end{align*}

\textbf{Case 4: Correct outcome model and wrong treatment model} \\
In Case 4, we assumed that users correctly specified the outcome model but misspecified the treatment model with $X_2$ omitted. 
\begin{align*}
h(t | Z, {\bf X}) &= h_0(t) \exp(\beta_0 {Z} + \beta_1 X_1 + \beta_2 X_2 + \beta_4 X_4 + \beta_5 X_5) \\
E[Z = 1 | {\bf X}] &= g(\alpha_0 + \alpha_3 X_3 + \alpha_5 X_5 + \alpha_6 X_6).
\end{align*}

\textbf{Case 5: Both models wrong by variable omission} \\
In Case 5, we assumed that users omitted $X_2$ from the outcome model and the treatment model. 
\begin{align*}
h(t | Z, {\bf X}) &= h_0(t) \exp(\beta_0 {Z} + \beta_1 X_1 + \beta_4 X_4 + \beta_5 X_5 )\\
E[Z = 1 | {\bf X}] &= g(\alpha_0 + \alpha_3 X_3 + \alpha_5 X_5 + \alpha_6 X_6).
\end{align*}

Note that, regarding model properties, neither a treatment model nor an outcome model is used for KM. For DS, only outcome model is used. For IPTW KM, IPTW CH, and Matching, only treatment model is used. For EL, neither a treatment model nor an outcome model is used, but the covariates specified in treatment model will be used to estimate the likelihood in our numerical studies. For AIPTW, both outcome and treatment models are used.

\subsubsection{Population Parameters of Interest and Statistical Inference} 
Our analyses focused on four key parameters to evaluate the AH  with a truncation time $\tau=0.7$: 
${\eta}_1(\tau)$, ${\eta}_0(\tau)$, 
${\eta}_1(\tau)-{\eta}_0(\tau)$, and 
${\eta}_1(\tau)/{\eta}_0(\tau)$, 
representing the AH in Group 0 and Group 1, DAH, and RAH, respectively. 
We estimated these parameters based on the adjusted survival curves, $\hat{S}_1(t)$ and $\hat{S}_0(t),$ derived through the methods described in Section 2.2.  
We also calculated 0.95 confidence intervals using bootstrap.
Specifically, for each dataset, we generated a bootstrap sample and estimated ${\eta}_1(\tau)$, ${\eta}_0(\tau)$, DAH, and RAH for each method. The process of generating a bootstrap sampling and estimating the parameters was repeated 300 times, enabling us to compute standard errors of the point estimators and corresponding 0.95 confidence intervals. Note that standard error and confidence intervals for RAH were calculated on the log-scale.

\subsection{Performance Evaluation}

\subsubsection{Derivation of True Parameter Values}
The true values of the four parameters (i.e., ${\eta}_1(\tau)$, ${\eta}_0(\tau)$, DAH, and RAH) under each scenario were derived numerically. 
Specifically, we randomly generated 10,000 data points of the covariate vectors $\bf{X}.$
We then generated the event time $T$ when $Z=1,$ using the formula in Section 2.5. 
Using the exact same $\bf{X},$ we generated $T$ when $Z=0$ in the same manner. 
Combining these two datasets for the treatment $(Z=1)$ and control $(Z=0)$ groups together, we derived an estimate for each of the four parameters using the standard KM approach. To assure that the Monte Carlo error affecting the true values we computed was negligible, we repeated the process 100 times and averaged the estimates for each parameter.

\subsubsection{Performance Metrics}

For each of these parameters, we employed a comprehensive set of metrics to assess the performance of various confounding adjustment methods when used with the AH-based approach in CER. 
Firstly, we calculated the relative bias of point estimates, providing a scaled perspective of the bias relative to the true parameter value. 
Secondly, we computed the square-root of the Mean Squared Error (rMSE) for the point estimates. This metric, by incorporating both bias and variance, provides a comprehensive overview of the overall estimation error.
Thirdly, we assessed the coverage probability of 0.95 confidence interval, which reflected the proportion of times the true parameter value falls within the calculated 0.95 confidence interval. 
Fourthly, the median length of 0.95 confidence interval, while being robust to the outliers, was calculated to offer insight into the efficiency of the estimating procedures. 

\section{Results}

We present the results from DAH (Tables 1 to 4), one of our four AH-based parameters, with the four performance metrics mentioned above when the total sample size is $n=300.$ 
The results for the remaining parameters are reported in the Appendix (see Tables A1 to A12).

\begin{table}[b]
\centering
\caption{Relative Bias for DAH}
\label{tab:DAH Relative Bias}
\begin{tabular}{cclccccc}
\hline
&  &  & & & Case  &  & \\
Scenario & Censoring & Method & 1 & 2 & 3 & 4 & 5 \\ \hline
1 & A & KM & -0.2766 & -0.2766 & -0.2766 & -0.2766 & -0.2766 \\ 
 &  & DS & -0.0299 & -0.0317 & -0.1608 & -0.0299 & -0.1608 \\ 
 &  & IPTW KM & -0.0068 & -0.0055 & -0.0068 & -0.1444 & -0.1444 \\ 
 &  & IPTW CH & 0.0142 & 0.0155 & 0.0142 & -0.1271 & -0.1271 \\ 
 &  & Matching & -0.0108 & -0.0065 & -0.0108 & -0.1420 & -0.1420 \\ 
 &  & EL & 0.0067 & -0.0047 & 0.0067 & -0.1581 & -0.1581 \\ 
 &  & AIPTW & -0.0083 & -0.0072 & -0.0094 & -0.0101 & -0.1455 \\ \hline
1 & B & KM & -0.2822 & -0.2822 & -0.2822 & -0.2822 & -0.2822 \\ 
 &  & DS & -0.0348 & -0.0365 & -0.1651 & -0.0348 & -0.1651 \\ 
 &  & IPTW KM & -0.0098 & -0.0089 & -0.0098 & -0.1477 & -0.1477 \\ 
 &  & IPTW CH & 0.0075 & 0.0085 & 0.0075 & -0.1332 & -0.1332 \\ 
 &  & Matching & -0.0146 & -0.0120 & -0.0146 & -0.1444 & -0.1444 \\ 
 &  & EL & 0.0011 & -0.0054 & 0.0011 & -0.1610 & -0.1610 \\ 
 &  & AIPTW & -0.0114 & -0.0110 & -0.0129 & -0.0143 & -0.1495 \\ \hline
2 & A & KM & -0.2636 & -0.2636 & -0.2636 & -0.2636 & -0.2636 \\ 
 &  & DS & -0.0420 & -0.0435 & -0.1679 & -0.0420 & -0.1679 \\ 
 &  & IPTW KM & 0.0011 & 0.0011 & 0.0011 & -0.1373 & -0.1373 \\ 
 &  & IPTW CH & -0.0055 & -0.0055 & -0.0055 & -0.1417 & -0.1417 \\ 
 &  & Matching & 0.0057 & 0.0020 & 0.0057 & -0.1368 & -0.1368 \\ 
 &  & EL & 0.0153 & 0.0041 & 0.0153 & -0.1472 & -0.1472 \\ 
 &  & AIPTW & -0.0003 & 0.0005 & -0.0013 & -0.0061 & -0.1380 \\ \hline
2 & B & KM & -0.2667 & -0.2667 & -0.2667 & -0.2667 & -0.2667 \\ 
 &  & DS & -0.0496 & -0.0517 & -0.1754 & -0.0496 & -0.1754 \\ 
 &  & IPTW KM & -0.0055 & -0.0054 & -0.0055 & -0.1421 & -0.1421 \\ 
 &  & IPTW CH & -0.0119 & -0.0119 & -0.0119 & -0.1466 & -0.1466 \\ 
 &  & Matching & -0.0080 & -0.0054 & -0.0080 & -0.1361 & -0.1361 \\ 
 &  & EL & 0.0068 & -0.0017 & 0.0068 & -0.1550 & -0.1550 \\ 
 &  & AIPTW & -0.0084 & -0.0082 & -0.0097 & -0.0125 & -0.1449 \\ \hline
\end{tabular}
\centering
\\
\vspace{0.3cm}
\begin{minipage}{360pt}
Methods for deriving survival curves: 
KM, Standard Kaplan-Meier based approach (unadjusted); 
DS, Direct Standardization via a Cox model (G-formula); 
IPTW KM, Xie and Liu's approach;
IPTW CH, Cole and Hernan's approach; 
Matching, Propensity score matching;
EL, Empirical Likelihood approach;
AIPTW, Augmented Inverse Probability of Treatment Weighting approach.\\
\begin{tabular}{cll}
Case & Outcome model & Treatment model \\ \hline  
1 & Correct & Correct \\
2 & Included extra variables, $X_3$ and $X_6$ & 
    Included extra variables, $X_1$ and $X_4$ \\
3 & Failed to include $X_2$ & Correct \\
4 & Correct & Failed to include $X_2$ \\
5 & Failed to include $X_2$ & Failed to include $X_2$ \\ \hline 
\end{tabular}
\end{minipage}
\end{table}

\begin{table}[b]
\centering
\caption{Square-root of Mean Square Error (rMSE) for DAH}
\label{tab:DAH RMSE}
\begin{tabular}{cclccccc}
\hline
&  &  & & & Case  &  & \\
Scenario & Censoring & Method & 1 & 2 & 3 & 4 & 5 \\ \hline
1 & A & KM & 0.2893 & 0.2893 & 0.2893 & 0.2893 & 0.2893 \\ 
 &  & DS & 0.1419 & 0.1509 & 0.1901 & 0.1419 & 0.1901 \\ 
 &  & IPTW KM & 0.2003 & 0.1895 & 0.2003 & 0.2263 & 0.2263 \\ 
 &  & IPTW CH & 0.2068 & 0.1961 & 0.2068 & 0.2250 & 0.2250 \\ 
 &  & Matching & 0.2370 & 0.2241 & 0.2370 & 0.2506 & 0.2506 \\ 
 &  & EL & 0.2659 & 0.2449 & 0.2659 & 0.2556 & 0.2556 \\ 
 &  & AIPTW & 0.1917 & 0.1931 & 0.1919 & 0.1842 & 0.2187 \\ \hline
1 & B & KM & 0.2954 & 0.2954 & 0.2954 & 0.2954 & 0.2954 \\ 
 &  & DS & 0.1463 & 0.1560 & 0.1952 & 0.1463 & 0.1952 \\ 
 &  & IPTW KM & 0.2071 & 0.1966 & 0.2071 & 0.2336 & 0.2336 \\ 
 &  & IPTW CH & 0.2121 & 0.2016 & 0.2121 & 0.2329 & 0.2329 \\ 
 &  & Matching & 0.2441 & 0.2310 & 0.2441 & 0.2572 & 0.2572 \\ 
 &  & EL & 0.2720 & 0.2390 & 0.2720 & 0.2574 & 0.2574 \\ 
 &  & AIPTW & 0.1977 & 0.1997 & 0.1980 & 0.1901 & 0.2254 \\ \hline
2 & A & KM & 0.5248 & 0.5248 & 0.5248 & 0.5248 & 0.5248 \\ 
 &  & DS & 0.2837 & 0.2970 & 0.3744 & 0.2837 & 0.3744 \\ 
 &  & IPTW KM & 0.3723 & 0.3486 & 0.3723 & 0.4141 & 0.4141 \\ 
 &  & IPTW CH & 0.3665 & 0.3429 & 0.3665 & 0.4134 & 0.4134 \\ 
 &  & Matching & 0.4273 & 0.4170 & 0.4273 & 0.4705 & 0.4705 \\ 
 &  & EL & 0.4362 & 0.4867 & 0.4362 & 0.4447 & 0.4447 \\ 
 &  & AIPTW & 0.3480 & 0.3497 & 0.3488 & 0.3364 & 0.3952 \\ \hline
2 & B & KM & 0.5315 & 0.5315 & 0.5315 & 0.5315 & 0.5315 \\ 
 &  & DS & 0.2911 & 0.3074 & 0.3874 & 0.2911 & 0.3874 \\ 
 &  & IPTW KM & 0.3828 & 0.3600 & 0.3828 & 0.4301 & 0.4301 \\ 
 &  & IPTW CH & 0.3765 & 0.3537 & 0.3765 & 0.4289 & 0.4289 \\ 
 &  & Matching & 0.4524 & 0.4253 & 0.4524 & 0.4731 & 0.4731 \\ 
 &  & EL & 0.5056 & 0.4302 & 0.5056 & 0.4680 & 0.4680 \\ 
 &  & AIPTW & 0.3627 & 0.3658 & 0.3633 & 0.3524 & 0.4155 \\ \hline
\end{tabular}
\centering
\\
\vspace{0.3cm}
\begin{minipage}{360pt}
Methods for deriving survival curves: 
KM, Standard Kaplan-Meier based approach (unadjusted); 
DS, Direct Standardization via a Cox model (G-formula); 
IPTW KM, Xie and Liu's approach;
IPTW CH, Cole and Hernan's approach; 
Matching, Propensity score matching;
EL, Empirical Likelihood approach;
AIPTW, Augmented Inverse Probability of Treatment Weighting approach.\\
\begin{tabular}{cll}
Case & Outcome model & Treatment model \\ \hline  
1 & Correct & Correct \\
2 & Included extra variables, $X_3$ and $X_6$ & 
    Included extra variables, $X_1$ and $X_4$ \\
3 & Failed to include $X_2$ & Correct \\
4 & Correct & Failed to include $X_2$ \\
5 & Failed to include $X_2$ & Failed to include $X_2$ \\ \hline 
\end{tabular}
\end{minipage}\end{table}

\begin{table}[b]
\centering
\caption{Coverage Probability of 0.95 Confidence Interval for DAH}
\label{tab:DAH Coverage Probability}
\begin{tabular}{cclccccc}
\hline
 &  &  & & & Case  &  & \\
Scenario & Censoring & Method & 1 & 2 & 3 & 4 & 5 \\ \hline
1 & A & KM & 0.815 & 0.815 & 0.815 & 0.815 & 0.815 \\ 
 &  & DS & 0.949 & 0.951 & 0.862 & 0.949 & 0.862 \\ 
 &  & IPTW KM & 0.955 & 0.955 & 0.955 & 0.924 & 0.924 \\ 
 &  & IPTW CH & 0.956 & 0.963 & 0.956 & 0.941 & 0.941 \\ 
 &  & Matching & 0.948 & 0.955 & 0.948 & 0.927 & 0.927 \\ 
 &  & EL & 0.962 & 0.973 & 0.962 & 0.924 & 0.924 \\ 
 &  & AIPTW & 0.976 & 0.976 & 0.977 & 0.976 & 0.956 \\ \hline
1 & B & KM & 0.810 & 0.810 & 0.810 & 0.810 & 0.810 \\ 
 &  & DS & 0.949 & 0.947 & 0.861 & 0.949 & 0.861 \\ 
 &  & IPTW KM & 0.952 & 0.955 & 0.952 & 0.921 & 0.921 \\ 
 &  & IPTW CH & 0.955 & 0.960 & 0.955 & 0.936 & 0.936 \\ 
 &  & Matching & 0.942 & 0.952 & 0.942 & 0.928 & 0.928 \\ 
 &  & EL & 0.958 & 0.974 & 0.958 & 0.923 & 0.923 \\ 
 &  & AIPTW & 0.984 & 0.983 & 0.983 & 0.987 & 0.964 \\ \hline
2 & A & KM & 0.771 & 0.771 & 0.771 & 0.771 & 0.771 \\ 
 &  & DS & 0.946 & 0.945 & 0.842 & 0.946 & 0.842 \\ 
 &  & IPTW KM & 0.957 & 0.961 & 0.957 & 0.920 & 0.920 \\ 
 &  & IPTW CH & 0.956 & 0.961 & 0.956 & 0.915 & 0.915 \\ 
 &  & Matching & 0.960 & 0.961 & 0.960 & 0.916 & 0.916 \\ 
 &  & EL & 0.966 & 0.971 & 0.966 & 0.924 & 0.924 \\ 
 &  & AIPTW & 0.973 & 0.975 & 0.974 & 0.974 & 0.940 \\ \hline
2 & B & KM & 0.771 & 0.771 & 0.771 & 0.771 & 0.771 \\ 
 &  & DS & 0.939 & 0.945 & 0.838 & 0.939 & 0.838 \\ 
 &  & IPTW KM & 0.951 & 0.960 & 0.951 & 0.906 & 0.906 \\ 
 &  & IPTW CH & 0.952 & 0.958 & 0.952 & 0.901 & 0.901 \\ 
 &  & Matching & 0.948 & 0.955 & 0.948 & 0.920 & 0.920 \\ 
 &  & EL & 0.956 & 0.976 & 0.956 & 0.909 & 0.909 \\ 
 &  & AIPTW & 0.980 & 0.979 & 0.981 & 0.981 & 0.947 \\ \hline
\end{tabular}
\centering
\\
\vspace{0.3cm}
\begin{minipage}{360pt}
Methods for deriving survival curves: 
KM, Standard Kaplan-Meier based approach (unadjusted); 
DS, Direct Standardization via a Cox model (G-formula); 
IPTW KM, Xie and Liu's approach;
IPTW CH, Cole and Hernan's approach; 
Matching, Propensity score matching;
EL, Empirical Likelihood approach;
AIPTW, Augmented Inverse Probability of Treatment Weighting approach.\\
\begin{tabular}{cll}
Case & Outcome model & Treatment model \\ \hline  
1 & Correct & Correct \\
2 & Included extra variables, $X_3$ and $X_6$ & 
    Included extra variables, $X_1$ and $X_4$ \\
3 & Failed to include $X_2$ & Correct \\
4 & Correct & Failed to include $X_2$ \\
5 & Failed to include $X_2$ & Failed to include $X_2$ \\ \hline 
\end{tabular}
\end{minipage}\end{table}

\begin{table}[b]
\centering
\caption{Median Length of 0.95 Confidence Interval for DAH}
\label{tab:DAH Median CI Length}
\begin{tabular}{cclccccc}
\hline
&  &  & & & Case  &  & \\
Scenario & Censoring & Method & 1 & 2 & 3 & 4 & 5 \\ \hline
1 & A & KM & 0.7677 & 0.7677 & 0.7677 & 0.7677 & 0.7677 \\ 
 &  & DS & 0.5543 & 0.5866 & 0.5590 & 0.5543 & 0.5590 \\ 
 &  & IPTW KM & 0.7893 & 0.7539 & 0.7893 & 0.7681 & 0.7681 \\ 
 &  & IPTW CH & 0.8308 & 0.7966 & 0.8308 & 0.8122 & 0.8122 \\ 
 &  & Matching & 0.9073 & 0.9057 & 0.9073 & 0.8836 & 0.8836 \\ 
 &  & EL & 0.9361 & 0.9883 & 0.9361 & 0.8250 & 0.8250 \\ 
 &  & AIPTW & 0.8246 & 0.8296 & 0.8270 & 0.8056 & 0.8071 \\ \hline
1 & B & KM & 0.7750 & 0.7750 & 0.7750 & 0.7750 & 0.7750 \\ 
 &  & DS & 0.5665 & 0.5980 & 0.5693 & 0.5665 & 0.5693 \\ 
 &  & IPTW KM & 0.8067 & 0.7710 & 0.8067 & 0.7824 & 0.7824 \\ 
 &  & IPTW CH & 0.8411 & 0.8044 & 0.8411 & 0.8196 & 0.8196 \\ 
 &  & Matching & 0.9260 & 0.9228 & 0.9260 & 0.8997 & 0.8997 \\ 
 &  & EL & 0.9498 & 1.0028 & 0.9498 & 0.8379 & 0.8379 \\ 
 &  & AIPTW & 0.9198 & 0.9268 & 0.9237 & 0.8994 & 0.8965 \\ \hline
2 & A & KM & 1.3432 & 1.3432 & 1.3432 & 1.3432 & 1.3432 \\ 
 &  & DS & 1.1001 & 1.1526 & 1.0946 & 1.1001 & 1.0946 \\ 
 &  & IPTW KM & 1.4761 & 1.4031 & 1.4761 & 1.4360 & 1.4360 \\ 
 &  & IPTW CH & 1.4517 & 1.3809 & 1.4517 & 1.4171 & 1.4171 \\ 
 &  & Matching & 1.6994 & 1.6868 & 1.6994 & 1.6477 & 1.6477 \\ 
 &  & EL & 1.7214 & 1.7796 & 1.7214 & 1.5268 & 1.5268 \\ 
 &  & AIPTW & 1.4933 & 1.4984 & 1.4953 & 1.4577 & 1.4454 \\ \hline
2 & B & KM & 1.3567 & 1.3567 & 1.3567 & 1.3567 & 1.3567 \\ 
 &  & DS & 1.1105 & 1.1662 & 1.1043 & 1.1105 & 1.1043 \\ 
 &  & IPTW KM & 1.5033 & 1.4293 & 1.5033 & 1.4523 & 1.4523 \\ 
 &  & IPTW CH & 1.4768 & 1.4052 & 1.4768 & 1.4343 & 1.4343 \\ 
 &  & Matching & 1.7231 & 1.7081 & 1.7231 & 1.6655 & 1.6655 \\ 
 &  & EL & 1.7386 & 1.7978 & 1.7386 & 1.5389 & 1.5389 \\ 
 &  & AIPTW & 1.6011 & 1.6146 & 1.5978 & 1.5624 & 1.5453 \\ \hline
\end{tabular}
\centering
\\
\vspace{0.3cm}
\begin{minipage}{360pt}
Methods for deriving survival curves: 
KM, Standard Kaplan-Meier based approach (unadjusted); 
DS, Direct Standardization via a Cox model (G-formula); 
IPTW KM, Xie and Liu's approach;
IPTW CH, Cole and Hernan's approach; 
Matching, Propensity score matching;
EL, Empirical Likelihood approach;
AIPTW, Augmented Inverse Probability of Treatment Weighting approach.\\
\begin{tabular}{cll}
Case & Outcome model & Treatment model \\ \hline  
1 & Correct & Correct \\
2 & Included extra variables, $X_3$ and $X_6$ & 
    Included extra variables, $X_1$ and $X_4$ \\
3 & Failed to include $X_2$ & Correct \\
4 & Correct & Failed to include $X_2$ \\
5 & Failed to include $X_2$ & Failed to include $X_2$ \\ \hline 
\end{tabular}
\end{minipage}\end{table}

Table \ref{tab:DAH Relative Bias} presents relative bias for DAH given different confounding adjustment methods under various scenarios, censoring patterns, and cases. 
KM provided a large bias (ranged from -0.28 to -0.26) in all situations, due to the lack of confounding adjustment.
The relative bias by DS was low (ranged from -0.05 to -0.03) when the outcome model was correctly specified (Case 1 and 4), but it was remarkable (ranged -0.18 to -0.16) when the outcome model failed to include a predictor (Case 3 and 5). For Case 2, where the outcome model included extra variables, the relative bias was almost identical to the ones in Case 1 and 4. This suggests that the relative bias by the DS method is more significantly affected by omitting covariates compared to adding extra covariates.
The relative bias produced by IPTW KM, IPTW CH, and Matching depended on the adequacy of the treatment model, as a treatment model is used for adjustment in these approaches. 
When the treatment is correct (Case 1 and 3), the resulting relative bias was small (ranged from -0.02 to 0.01). 
The relative bias with Case 2, where two extra variables were included in the treatment model, was similar to the results with Case 1 and 3. In contrast, when a prognostic factor $X_2$ was omitted from the treatment model (Case 4 and 5), large relative biases (ranged from -0.15 to -0.13) were observed. 
In EL, the covariates included in the treatment model were used for adjustment. 
Thus, Case 1 and 3 were the situations where the set of variables were associated with the treatment assignment. 
In the case where two extra variables were included (Case 2), the absolute value of the relative bias produced by EL was less than 0.01. Where EL failed to include a variable that was associated with the treatment assignment (Case 4 and 5), large relative biases (ranged from -0.16 to -0.15) were observed. 
As AIPTW has the doubly robust property, theoretically, it would produce an unbiased estimate when either the outcome model or the treatment model is correctly specified (Case 1, 3, and 4). 
As expected, the absolute values of the relative bias in these cases were at most 0.014 in our simulation scenarios.  
Similar to the other methods, Case 2 produced similar results. 
However, when neither model was correct (Case 5), AIPTW produced large relative biases (ranged from -0.15 to -0.14).

Table \ref{tab:DAH RMSE} presents the results of the rMSE for DAH. Analogous to the findings related to relative bias, the rMSE was significantly influenced by the appropriateness of the specified model(s) for each method. For instance, the DS method relies on the outcome model. The rMSE values were lower in cases where the outcome model was correctly specified (Case 1 or 4) or included all necessary predictors (Case 2), compared to cases where the outcome model omitted a crucial predictor. 
In Case 1, where both the outcome and treatment models were correct, the DS method provided the lowest rMSE compared to other methods. Specifically, under Scenario 1 with Censoring Pattern A, the rMSE of DS was 0.142, while the rMSE of the other adjusted methods ranged from 0.192 to 0.266. The rMSE of Matching (0.237) was larger than the other propensity score adjustment methods (0.200 with IPTW KM and 0.207 with IPTW CH). The EL method provided the largest rMSE (0.266) among these adjusted approaches. 

Table \ref{tab:DAH Coverage Probability} presents coverage probability of 95\% confidence interval for DAH given different confounding adjustment methods under various scenarios, censoring patterns, and cases. 
KM provided lower coverage probabilities (ranged from 0.771 to 0.815) than the nominal level in all situations, due to the lack of confounding adjustment.
DS had low coverage probabilities under Case 3 and 5, where the outcome model failed to include a predictor. Otherwise, the coverage probabilities of DS were closer to the nominal level.  
The coverage probabilities of IPTW CH, IPTW KM, and Matching were also off from the nominal level when the treatment model did not include a crucial variable (Case 4 and 5). 
The coverage probabilities of EL ranged from 0.956 to 0.966, which were close to the nominal level, when a correct set of variables were specified for the adjustment (Case 1 and 3). When extra variables were included for adjustment (Case 2), the coverage probabilities were higher (0.97 to  to 0.98) than the nominal level. When a crucial variable was not included for adjustment (Case 4 and 5), EL had low coverage probabilities (0.91 to 0.92).
Theoretically, AIPTW works well under Case 1, 3, and 4 due to its doubly robust property. 
However, the coverage probabilities of AIPTW tended to be conservative in these cases. 
For example, in Case 1, the coverage probabilities ranged from 0.973 and 0.984, which were higher than the nominal 0.95 level.

Table \ref{tab:DAH Median CI Length} presents median length of the confidence interval for DAH.
Overall, the median length of DS was shorter than those of the other methods. 
For example, in Case 1 under Scenario 2 with Censoring Pattern A, 
the median length of DS was 1.10, while the lengths of the other adjusting methods ranged from 1.45 to 1.72. 
The Matching and EL methods produced relatively wider confidence intervals compared to the other adjusting methods. For example, in Case 1 under Scenario 2 with Censoring Pattern B, 
the median lengths of Matching and EL were 1.72 and 1.74, respectively, while the lengths of the other adjusting methods ranged from 1.11 to 1.60.

In summary, the results of this simulation study suggest two primary insights. First, when there is confidence in the correct specification of the outcome model, DS emerges as the optimal method for confounding adjustment due to its consistently stable performance across our experiments. If the outcome model's specification is uncertain, then AIPTW is generally recommended. AIPTW is preferred over approaches that only use either a treatment model or an outcome model due to its doubly robust property, which is particularly advantageous given the similar efficiency reflected by the median length of confidence intervals. Second, despite the theoretical expectations of EL due to its robust property, it did not perform well within the framework of our numerical studies. We found that across all assessed performance measures, AIPTW was generally superior to EL under our simulation settings.

There are two points that should be noted regarding numerical issues we experienced in obtaining the adjusted survival curves using the existing R program packages. 
Firstly, for EL, we encountered a convergence warning several times during the bootstrap process, although the R package still returned the numbers. This might explain why the performance of EL was not as expected. Secondly, regarding the AIPTW method, certain bootstrapped datasets failed due to an "invalid pointer" error stemming from a C++ issue related to matrix calculations. To counteract this, we increased the number of iterations for the simulation and discarded unsuccessful iterations, ensuring 2000 iterations to calculate the performance measures.

\section{Use Case Example}

To illustrate the methods explored in this study, we utilized data from the pragmatic effectiveness CANVAS trial (NCT02744092; AFT-28) comparing DOACs and LMWH in preventing recurrent VTE among cancer patients with an initial VTE event. This study was conducted across 67 oncology practices in the US, and has a hybrid design with a randomization cohort and a preference cohort. 
A total of 671 patients consented to randomization and were randomized to receive either DOACs or LMWH. 
Another 140 patients declined randomization and were placed in the preference cohort where they chose either DOACs or LMWH. Of the 140, three patients did not receive the selected treatment. 
The primary outcome was recurrent VTE. Major bleeding and death were secondary outcomes. The detailed results with the randomization cohort data have been reported in \cite{Schrag2023-ym}.

\begin{figure}[b]
  \centering
  \includegraphics[width=0.8\textwidth]{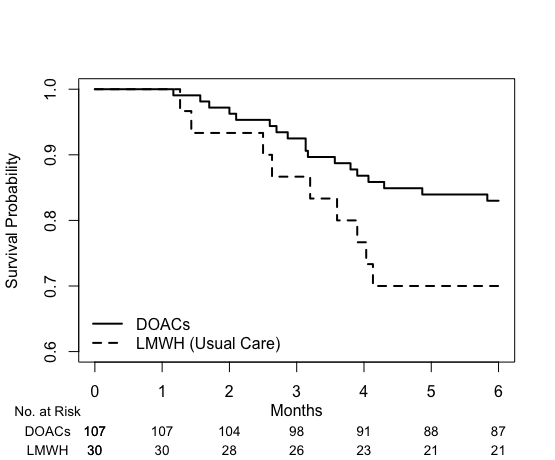}
  \caption{Estimated Survival Curves for DOAC (solid line) and LMWH (dashed line) using the Kaplan-Meier Method}
  \label{fig: Unadjusted figures}
\end{figure}

In this paper, we used the overall survival data from the 137 patients who received the selected treatment in the preference cohort. Figure \ref{fig: Unadjusted figures} shows the estimated survival curves for the DOAC and LMWH groups based on the KM method. The estimated survival probabilities for the DOAC group were uniformly higher than those of the LMWH group throughout the observation period. The AH values for the DOAC and LMWH groups were 0.039 and 0.079, respectively. In the unadjusted analysis, while the DOAC group seemed to perform better than the LMWH group, this difference was not statistically significant. Specifically, the estimated difference in AH was -0.039, with a corresponding 0.95 confidence interval ranging from -0.103 to 0.024, which included zero.

Due to the lack of randomization in the treatment allocation for patients within the preference cohort, the unadjusted analysis was potentially influenced by selection bias. Thus, we performed the adjustment methods outlined in previous sections. For the execution of these adjusted analyses, we derived an outcome model and a treatment model. The details of these models are presented in Table \ref{tab:outcome model} for the outcome model and Table \ref{tab:ps model} for the treatment model, respectively.

Table \ref{tab:canvas example} presents both unadjusted and adjusted AH with a truncation time of 6 months. 
For example, the AH values adjusted by the AIPTW method were 0.032 for the DOAC group and 0.046 for the LMWH group. Difference in the adjusted AH was -0.014 (0.95 confidence interval: -0.048 to 0.020).
All methods for adjusting confounders yielded numerically similar results. 
These approaches uniformly indicated that, in the preference cohort, after adjustment for confounding factors, DOAC does not offer a significant survival benefit over LMWH. This outcome aligns with the observations from the randomized cohort study.

\begin{table}[b]
\centering
\caption{Outcome Model: Cox Regression}
\label{tab:outcome model}
\begin{tabular}{lccc}
\hline
Factor & Hazard Ratio & 0.95 CI & p-value \\ \hline
Treatment (DOACs vs LMWH) & 0.70 & (0.31, 1.60) & 0.397 \\ 
Sex (Female vs Male) & 2.95 & (1.21, 7.19) & 0.017 \\ 
Albumin (high vs low)& 0.23 & (0.09, 0.56) &  0.001 \\ 
Metastases (Yes vs No) & 2.97 & (0.85, 10.38) & 0.088 \\ 
Pulmonary Embolism (Yes vs No) & 0.58 & (0.26, 1.31) & 0.190 \\  \hline
\end{tabular}
\end{table}

\begin{table}[b]
\centering
\caption{Treatment Model (Propensity Score Model): Logistic Regression}
\label{tab:ps model}
\begin{tabular}{lccc}
\hline
Factor & Odds Ratio & 0.95 CI & p-value \\ \hline
(Intercept) & 6.10 & (2.48, 14.98) & $<$ 0.001 \\ 
Sex (Female vs Male) & 0.50 & (0.20, 1.24) & 0.134 \\ 
Indwelling central venous (Yes vs No) & 3.32 & (1.22, 9.03) &  0.019 \\ 
Metastases (Yes vs No) & 0.39 & (0.14, 1.12) & 0.080 \\ 
Bevacizumab (Yes vs No) & 0.27 & (0.08, 0.94) & 0.039 \\  \hline
\end{tabular}
\end{table}

\begin{table}[b]
\centering
\caption{Unadjusted and adjusted AH ($\tau = 6$ months) with the CANVAS mortality data from the preference cohort}
\label{tab:canvas example}
\begin{tabular}{lclccccc}
\hline
Method & DOAC & LMWH & Difference (0.95CI) & Ratio (0.95CI) \\ \hline
DS & 0.034 & 0.046 & -0.012 (-0.041 to 0.018) & 0.741 (0.345 to 1.593) \\ 
IPTW KM & 0.030 & 0.049 & -0.019 (-0.056 to 0.019) & 0.621 (0.239 to 1.614)  \\ 
IPTW CH & 0.038 & 0.063 & -0.025 (-0.080 to 0.031) & 0.610 (0.229 to 1.626) \\ 
Matching & 0.031 & 0.057 & -0.027 (-0.073 to 0.020) & 0.536 (0.184 to 1.567) \\ 
EL & 0.031 & 0.056 & -0.025 (-0.066 to 0.016) & 0.549 (0.122 to 2.462) \\ 
AIPTW & 0.032 & 0.046 & -0.014 (-0.048 to 0.020) & 0.701 (0.275 to 1.786)  \\ 
KM (Unadjusted) & 0.039 & 0.079 & -0.039 (-0.103 to 0.024) & 0.498 (0.194 to 1.281) \\
\hline
\end{tabular}
\end{table}

\section{Discussion}

We confirmed that existing methods to adjust survival function reasonably worked with the AH method in CER. If the outcome model is correct, the DS method will give unbiased results. If the treatment model is correct, methods based on propensity score will give unbiased results. If either outcome or treatment model is correct, AIPTW has decent performance. This conclusion is consistent with those obtained by \cite{denz2023comparison} for other summary measures defined as a function of the survival function, such as RMST or t-year event rate. 
Median survival time is also a robust summary metric of event time distribution, and difference and ratio of median could be used for summarizing the magnitude of the treatment effect. Although assessing the median survival time using confounding adjustment approaches was beyond our scope, we believe the confounding adjustment approach through the adjusted survival curves would also work well for median survival time. 

In our simulation studies, we employed Cox regression models for the outcome model. 
However, other models that can provide adjusted survival curves for each group, such as accelerated failure time models or proportional odds models, can also be used. 
While this paper focused on methods using adjusted survival curves, if the interest lies solely in a specific metric, a regression analysis for that metric can be employed. For instance, if the causal estimand is an RMST-based metric, a regression analysis specific to RMST (\cite{Tian2014-lj}) can be used, although this does not provide the adjusted survival curve. Similarly, if the estimand is an AH-based metric, as in our case, the regression analysis for AH recently proposed by \cite{Uno2024-lg} can also be a viable option.

In this numerical study, we solely utilized bootstrapping to derive standard errors for the construction of confidence intervals for all methods. Derivation of analytic variance estimates associated with AH for each confounding adjustment method is a promising direction for future investigations. Such an analytic framework could substantially decrease computational time, enabling the execution of experiments across a broader array of scenarios.

\section{Conclusions}

Our study has successfully demonstrated the feasibility of applying commonly used confounding adjustment methods to AH, with all methods delivering satisfactory performance as expected. Importantly, our findings underscore the robustness of the AIPTW method, attributed to its doubly-robust property, making it a recommended approach for future research. However, we also identified limitations within the EL method. Despite its theoretical benefits, EL was characterized by wider confidence intervals and relatively unstable point estimates, suggesting the need for further research before its practical application can be endorsed for AH.

\section*{Supporting information}
Funding/Support: AFT-28 was funded through a Patient-Centered Outcomes Research Institute (PCORI) Award (CE-1304-6543). All statements in this publication, including its findings, are solely those of the authors and do not necessarily represent the views of the Patient-Centered Outcomes Research Institute (PCORI), its Board of Governors or Methodology Committee.

Role of the Funder/Sponsor: PCORI, the study funder, oversaw research ethics for the study, but had no role in the design and conduct of the study; collection, management, analysis, and interpretation of the data; preparation, review, or approval of the manuscript; or decision to submit the manuscript for publication. The Foundation of the Alliance for Clinical Trials in Oncology, the study sponsor, was responsible for the conduct and management of the study. 

Research reported in this publication was also supported by the National Institute of General Medical Sciences of the National Institutes of Health under award number R01GM152499.

\bibliographystyle{biom}
\bibliography{refs}

\clearpage

\appendix
\section*{Appendix}
The results from Group 0 AH, Group 1 AH, and RAH are presented below. 
\begin{table}[ht]
\renewcommand\thetable{A1}
\centering
\caption{Relative Bias for AH in Group 0}
\label{tab:Group0 AH Relative Bias}
\begin{tabular}{cclccccc}
\hline
&  &  & & & Case  &  & \\
Scenario & Censoring & Method & 1 & 2 & 3 & 4 & 5 \\ \hline
1 & A & KM & -0.0404 & -0.0404 & -0.0404 & -0.0404 & -0.0404 \\ 
 &  & DS & -0.0044 & -0.0049 & -0.0339 & -0.0044 & -0.0339 \\ 
 &  & IPTW KM & -0.0012 & -0.0012 & -0.0012 & -0.0335 & -0.0335 \\ 
 &  & IPTW CH & 0.0189 & 0.0188 & 0.0189 & -0.0139 & -0.0139 \\ 
 &  & Matching & -0.0030 & -0.0010 & -0.0030 & -0.0328 & -0.0328 \\ 
 &  & EL & -0.0011 & -0.0054 & -0.0011 & -0.0365 & -0.0365 \\ 
 &  & AIPTW & -0.0021 & -0.0021 & -0.0026 & -0.0028 & -0.0344 \\ \hline
1 & B & KM & -0.0443 & -0.0443 & -0.0443 & -0.0443 & -0.0443 \\ 
 &  & DS & -0.0058 & -0.0064 & -0.0352 & -0.0058 & -0.0352 \\ 
 &  & IPTW KM & -0.0015 & -0.0016 & -0.0015 & -0.0336 & -0.0336 \\ 
 &  & IPTW CH & 0.0146 & 0.0144 & 0.0146 & -0.0177 & -0.0177 \\ 
 &  & Matching & -0.0032 & -0.0021 & -0.0032 & -0.0326 & -0.0326 \\ 
 &  & EL & -0.0013 & -0.0043 & -0.0013 & -0.0365 & -0.0365 \\ 
 &  & AIPTW & -0.0030 & -0.0030 & -0.0034 & -0.0036 & -0.0350 \\ \hline
2 & A & KM & -0.0696 & -0.0696 & -0.0696 & -0.0696 & -0.0696 \\ 
 &  & DS & -0.0102 & -0.0108 & -0.0463 & -0.0102 & -0.0463 \\ 
 &  & IPTW KM & 0.0037 & 0.0035 & 0.0037 & -0.0376 & -0.0376 \\ 
 &  & IPTW CH & -0.0042 & -0.0045 & -0.0042 & -0.0443 & -0.0443 \\ 
 &  & Matching & 0.0068 & 0.0066 & 0.0068 & -0.0377 & -0.0377 \\ 
 &  & EL & 0.0060 & 0.0026 & 0.0060 & -0.0403 & -0.0403 \\ 
 &  & AIPTW & 0.0026 & 0.0030 & 0.0021 & 0.0002 & -0.0387 \\ \hline
2 & B & KM & -0.0725 & -0.0725 & -0.0725 & -0.0725 & -0.0725 \\ 
 &  & DS & -0.0146 & -0.0153 & -0.0506 & -0.0146 & -0.0506 \\ 
 &  & IPTW KM & 0.0014 & 0.0010 & 0.0014 & -0.0393 & -0.0393 \\ 
 &  & IPTW CH & -0.0067 & -0.0073 & -0.0067 & -0.0463 & -0.0463 \\ 
 &  & Matching & 0.0011 & 0.0019 & 0.0011 & -0.0367 & -0.0367 \\ 
 &  & EL & 0.0030 & -0.0013 & 0.0030 & -0.0425 & -0.0425 \\ 
 &  & AIPTW & -0.0004 & -0.0004 & -0.0009 & -0.0020 & -0.0410 \\ \hline
\end{tabular}
\centering
\\
\vspace{0.3cm}
\begin{minipage}{360pt}
Methods for deriving survival curves: 
KM, Standard Kaplan-Meier based approach (unadjusted); 
DS, Direct Standardization via a Cox model (G-formula); 
IPTW KM, Xie and Liu's approach;
IPTW CH, Cole and Hernan's approach; 
Matching, Propensity score matching;
EL, Empirical Likelihood approach;
AIPTW, Augmented Inverse Probability of Treatment Weighting approach.\\
\begin{tabular}{cll}
Case & Outcome model & Treatment model \\ \hline  
1 & Correct & Correct \\
2 & Included extra variables, $X_3$ and $X_6$ & 
    Included extra variables, $X_1$ and $X_4$ \\
3 & Failed to include $X_2$ & Correct \\
4 & Correct & Failed to include $X_2$ \\
5 & Failed to include $X_2$ & Failed to include $X_2$ \\ \hline 
\end{tabular}
\end{minipage}\end{table}

\begin{table}[t!]
\renewcommand\thetable{A2}
\centering
\caption{Square-root of Mean Square Error (rMSE) for AH in Group 0}
\label{tab:Group0 AH RMSE}
\begin{tabular}{cclccccc}
\hline
&  &  & & & Case  &  & \\
Scenario & Censoring & Method & 1 & 2 & 3 & 4 & 5 \\ \hline
1 & A & KM & 0.1496 & 0.1496 & 0.1496 & 0.1496 & 0.1496 \\ 
 &  & DS & 0.1175 & 0.1203 & 0.1299 & 0.1175 & 0.1299 \\ 
 &  & IPTW KM & 0.1424 & 0.1392 & 0.1424 & 0.1475 & 0.1475 \\ 
 &  & IPTW CH & 0.1522 & 0.1490 & 0.1522 & 0.1437 & 0.1437 \\ 
 &  & Matching & 0.1667 & 0.1620 & 0.1667 & 0.1665 & 0.1665 \\ 
 &  & EL & 0.1971 & 0.1866 & 0.1971 & 0.1625 & 0.1625 \\ 
 &  & AIPTW & 0.1319 & 0.1324 & 0.1329 & 0.1299 & 0.1413 \\ \hline
1 & B & KM & 0.1534 & 0.1534 & 0.1534 & 0.1534 & 0.1534 \\ 
 &  & DS & 0.1191 & 0.1224 & 0.1321 & 0.1191 & 0.1321 \\ 
 &  & IPTW KM & 0.1442 & 0.1413 & 0.1442 & 0.1490 & 0.1490 \\ 
 &  & IPTW CH & 0.1513 & 0.1482 & 0.1513 & 0.1457 & 0.1457 \\ 
 &  & Matching & 0.1693 & 0.1636 & 0.1693 & 0.1685 & 0.1685 \\ 
 &  & EL & 0.1998 & 0.1795 & 0.1998 & 0.1614 & 0.1614 \\ 
 &  & AIPTW & 0.1337 & 0.1345 & 0.1345 & 0.1319 & 0.1435 \\ \hline
2 & A & KM & 0.3311 & 0.3311 & 0.3311 & 0.3311 & 0.3311 \\ 
 &  & DS & 0.2519 & 0.2563 & 0.2764 & 0.2519 & 0.2764 \\ 
 &  & IPTW KM & 0.3048 & 0.2949 & 0.3048 & 0.3051 & 0.3051 \\ 
 &  & IPTW CH & 0.3006 & 0.2908 & 0.3006 & 0.3094 & 0.3094 \\ 
 &  & Matching & 0.3444 & 0.3440 & 0.3444 & 0.3527 & 0.3527 \\ 
 &  & EL & 0.3394 & 0.4158 & 0.3394 & 0.3189 & 0.3189 \\ 
 &  & AIPTW & 0.2833 & 0.2839 & 0.2842 & 0.2777 & 0.2902 \\ \hline
2 & B & KM & 0.3363 & 0.3363 & 0.3363 & 0.3363 & 0.3363 \\ 
 &  & DS & 0.2529 & 0.2591 & 0.2836 & 0.2529 & 0.2836 \\ 
 &  & IPTW KM & 0.3003 & 0.2933 & 0.3003 & 0.3062 & 0.3062 \\ 
 &  & IPTW CH & 0.2965 & 0.2895 & 0.2965 & 0.3110 & 0.3110 \\ 
 &  & Matching & 0.3503 & 0.3422 & 0.3503 & 0.3452 & 0.3452 \\ 
 &  & EL & 0.4123 & 0.3553 & 0.4123 & 0.3244 & 0.3244 \\ 
 &  & AIPTW & 0.2801 & 0.2813 & 0.2817 & 0.2780 & 0.2948 \\ \hline
\end{tabular}
\centering
\\
\vspace{0.3cm}
\begin{minipage}{360pt}
Methods for deriving survival curves: 
KM, Standard Kaplan-Meier based approach (unadjusted); 
DS, Direct Standardization via a Cox model (G-formula); 
IPTW KM, Xie and Liu's approach;
IPTW CH, Cole and Hernan's approach; 
Matching, Propensity score matching;
EL, Empirical Likelihood approach;
AIPTW, Augmented Inverse Probability of Treatment Weighting approach.\\
\begin{tabular}{cll}
Case & Outcome model & Treatment model \\ \hline  
1 & Correct & Correct \\
2 & Included extra variables, $X_3$ and $X_6$ & 
    Included extra variables, $X_1$ and $X_4$ \\
3 & Failed to include $X_2$ & Correct \\
4 & Correct & Failed to include $X_2$ \\
5 & Failed to include $X_2$ & Failed to include $X_2$ \\ \hline 
\end{tabular}
\end{minipage}\end{table}

\begin{table}[h]
\renewcommand\thetable{A3}
\centering
\caption{Coverage Probability of 0.95 Confidence Interval for AH in Group 0}
\label{tab:Group0 AH Coverage Probability}
\begin{tabular}{cclccccc}
\hline
&  &  & & & Case  &  & \\
Scenario & Censoring & Method & 1 & 2 & 3 & 4 & 5 \\ \hline
1 & A & KM & 0.922 & 0.922 & 0.922 & 0.922 & 0.922 \\ 
 &  & DS & 0.959 & 0.957 & 0.918 & 0.959 & 0.918 \\ 
 &  & IPTW KM & 0.959 & 0.964 & 0.959 & 0.928 & 0.928 \\ 
 &  & IPTW CH & 0.961 & 0.965 & 0.961 & 0.951 & 0.951 \\ 
 &  & Matching & 0.947 & 0.957 & 0.947 & 0.929 & 0.929 \\ 
 &  & EL & 0.970 & 0.978 & 0.970 & 0.936 & 0.936 \\ 
 &  & AIPTW & 0.977 & 0.979 & 0.975 & 0.976 & 0.944 \\ \hline
1 & B & KM & 0.922 & 0.922 & 0.922 & 0.922 & 0.922 \\ 
 &  & DS & 0.961 & 0.959 & 0.919 & 0.961 & 0.919 \\ 
 &  & IPTW KM & 0.964 & 0.963 & 0.964 & 0.930 & 0.930 \\ 
 &  & IPTW CH & 0.965 & 0.966 & 0.965 & 0.952 & 0.952 \\ 
 &  & Matching & 0.946 & 0.955 & 0.946 & 0.930 & 0.930 \\ 
 &  & EL & 0.970 & 0.980 & 0.970 & 0.939 & 0.939 \\ 
 &  & AIPTW & 0.982 & 0.982 & 0.979 & 0.981 & 0.961 \\ \hline
2 & A & KM & 0.847 & 0.847 & 0.847 & 0.847 & 0.847 \\ 
 &  & DS & 0.949 & 0.948 & 0.898 & 0.949 & 0.898 \\ 
 &  & IPTW KM & 0.950 & 0.957 & 0.950 & 0.919 & 0.919 \\ 
 &  & IPTW CH & 0.944 & 0.952 & 0.944 & 0.909 & 0.909 \\ 
 &  & Matching & 0.958 & 0.954 & 0.958 & 0.914 & 0.914 \\ 
 &  & EL & 0.964 & 0.973 & 0.964 & 0.927 & 0.927 \\ 
 &  & AIPTW & 0.969 & 0.970 & 0.970 & 0.971 & 0.932 \\ \hline
2 & B & KM & 0.851 & 0.851 & 0.851 & 0.851 & 0.851 \\ 
 &  & DS & 0.946 & 0.946 & 0.890 & 0.946 & 0.890 \\ 
 &  & IPTW KM & 0.958 & 0.957 & 0.958 & 0.923 & 0.923 \\ 
 &  & IPTW CH & 0.950 & 0.950 & 0.950 & 0.914 & 0.914 \\ 
 &  & Matching & 0.949 & 0.954 & 0.949 & 0.927 & 0.927 \\ 
 &  & EL & 0.962 & 0.976 & 0.962 & 0.926 & 0.926 \\ 
 &  & AIPTW & 0.975 & 0.975 & 0.972 & 0.973 & 0.940 \\ \hline
\end{tabular}
\centering
\\
\vspace{0.3cm}
\begin{minipage}{360pt}
Methods for deriving survival curves: 
KM, Standard Kaplan-Meier based approach (unadjusted); 
DS, Direct Standardization via a Cox model (G-formula); 
IPTW KM, Xie and Liu's approach;
IPTW CH, Cole and Hernan's approach; 
Matching, Propensity score matching;
EL, Empirical Likelihood approach;
AIPTW, Augmented Inverse Probability of Treatment Weighting approach.\\
\begin{tabular}{cll}
Case & Outcome model & Treatment model \\ \hline  
1 & Correct & Correct \\
2 & Included extra variables, $X_3$ and $X_6$ & 
    Included extra variables, $X_1$ and $X_4$ \\
3 & Failed to include $X_2$ & Correct \\
4 & Correct & Failed to include $X_2$ \\
5 & Failed to include $X_2$ & Failed to include $X_2$ \\ \hline 
\end{tabular}
\end{minipage}\end{table}

\begin{table}[t!]
\renewcommand\thetable{A4}
\centering
\caption{Median Length of 0.95 Confidence Interval for AH in Group 0}
\label{tab:Group0 AH Median CI Length}
\begin{tabular}{cclccccc}
\hline
&  &  & & & Case  &  & \\
Scenario & Censoring & Method & 1 & 2 & 3 & 4 & 5 \\ \hline
1 & A & KM & 0.5476 & 0.5476 & 0.5476 & 0.5476 & 0.5476 \\ 
 &  & DS & 0.4831 & 0.4929 & 0.4754 & 0.4831 & 0.4754 \\ 
 &  & IPTW KM & 0.5730 & 0.5616 & 0.5730 & 0.5517 & 0.5517 \\ 
 &  & IPTW CH & 0.6096 & 0.5972 & 0.6096 & 0.5867 & 0.5867 \\ 
 &  & Matching & 0.6488 & 0.6501 & 0.6488 & 0.6261 & 0.6261 \\ 
 &  & EL & 0.6772 & 0.7812 & 0.6772 & 0.5775 & 0.5775 \\ 
 &  & AIPTW & 0.5907 & 0.5939 & 0.5859 & 0.5821 & 0.5699 \\ \hline
1 & B & KM & 0.5520 & 0.5520 & 0.5520 & 0.5520 & 0.5520 \\ 
 &  & DS & 0.4899 & 0.5004 & 0.4826 & 0.4899 & 0.4826 \\ 
 &  & IPTW KM & 0.5819 & 0.5708 & 0.5819 & 0.5597 & 0.5597 \\ 
 &  & IPTW CH & 0.6149 & 0.6027 & 0.6149 & 0.5909 & 0.5909 \\ 
 &  & Matching & 0.6592 & 0.6604 & 0.6592 & 0.6365 & 0.6365 \\ 
 &  & EL & 0.6876 & 0.7874 & 0.6876 & 0.5859 & 0.5859 \\ 
 &  & AIPTW & 0.6523 & 0.6509 & 0.6388 & 0.6349 & 0.6210 \\ \hline
2 & A & KM & 1.0552 & 1.0552 & 1.0552 & 1.0552 & 1.0552 \\ 
 &  & DS & 1.0112 & 1.0302 & 0.9811 & 1.0112 & 0.9811 \\ 
 &  & IPTW KM & 1.2018 & 1.1728 & 1.2018 & 1.1427 & 1.1427 \\ 
 &  & IPTW CH & 1.1856 & 1.1547 & 1.1856 & 1.1306 & 1.1306 \\ 
 &  & Matching & 1.3606 & 1.3542 & 1.3606 & 1.2996 & 1.2996 \\ 
 &  & EL & 1.3626 & 1.5196 & 1.3626 & 1.1935 & 1.1935 \\ 
 &  & AIPTW & 1.1982 & 1.1997 & 1.1953 & 1.1870 & 1.1441 \\ \hline
2 & B & KM & 1.0612 & 1.0612 & 1.0612 & 1.0612 & 1.0612 \\ 
 &  & DS & 1.0154 & 1.0379 & 0.9844 & 1.0154 & 0.9844 \\ 
 &  & IPTW KM & 1.2127 & 1.1820 & 1.2127 & 1.1486 & 1.1486 \\ 
 &  & IPTW CH & 1.1943 & 1.1645 & 1.1943 & 1.1371 & 1.1371 \\ 
 &  & Matching & 1.3618 & 1.3670 & 1.3618 & 1.3078 & 1.3078 \\ 
 &  & EL & 1.3836 & 1.5283 & 1.3836 & 1.1989 & 1.1989 \\ 
 &  & AIPTW & 1.2630 & 1.2644 & 1.2541 & 1.2513 & 1.2040 \\ \hline
\end{tabular}
\centering
\\
\vspace{0.3cm}
\begin{minipage}{360pt}
Methods for deriving survival curves: 
KM, Standard Kaplan-Meier based approach (unadjusted); 
DS, Direct Standardization via a Cox model (G-formula); 
IPTW KM, Xie and Liu's approach;
IPTW CH, Cole and Hernan's approach; 
Matching, Propensity score matching;
EL, Empirical Likelihood approach;
AIPTW, Augmented Inverse Probability of Treatment Weighting approach.\\
\begin{tabular}{cll}
Case & Outcome model & Treatment model \\ \hline  
1 & Correct & Correct \\
2 & Included extra variables, $X_3$ and $X_6$ & 
    Included extra variables, $X_1$ and $X_4$ \\
3 & Failed to include $X_2$ & Correct \\
4 & Correct & Failed to include $X_2$ \\
5 & Failed to include $X_2$ & Failed to include $X_2$ \\ \hline 
\end{tabular}
\end{minipage}\end{table}

\begin{table}[t!]
\renewcommand\thetable{A5}
\centering
\caption{Relative Bias for AH in Group 1}
\label{tab:Group 1 AH Relative Bias}
\begin{tabular}{cclccccc}
\hline
&  &  & & & Case  &  & \\
Scenario & Censoring & Method & 1 & 2 & 3 & 4 & 5 \\ \hline
1 & A & KM & 0.1606 & 0.1606 & 0.1606 & 0.1606 & 0.1606 \\ 
 &  & DS & 0.0174 & 0.0179 & 0.0741 & 0.0174 & 0.0741 \\ 
 &  & IPTW KM & 0.0037 & 0.0024 & 0.0037 & 0.0609 & 0.0609 \\ 
 &  & IPTW CH & 0.0229 & 0.0215 & 0.0229 & 0.0825 & 0.0825 \\ 
 &  & Matching & 0.0036 & 0.0038 & 0.0036 & 0.0601 & 0.0601 \\ 
 &  & EL & -0.0078 & -0.0059 & -0.0078 & 0.0669 & 0.0669 \\ 
 &  & AIPTW & 0.0031 & 0.0022 & 0.0033 & 0.0034 & 0.0602 \\ \hline
1 & B & KM & 0.1581 & 0.1581 & 0.1581 & 0.1581 & 0.1581 \\ 
 &  & DS & 0.0189 & 0.0193 & 0.0754 & 0.0189 & 0.0754 \\ 
 &  & IPTW KM & 0.0056 & 0.0045 & 0.0056 & 0.0636 & 0.0636 \\ 
 &  & IPTW CH & 0.0206 & 0.0194 & 0.0206 & 0.0807 & 0.0807 \\ 
 &  & Matching & 0.0065 & 0.0063 & 0.0065 & 0.0626 & 0.0626 \\ 
 &  & EL & -0.0034 & -0.0035 & -0.0034 & 0.0695 & 0.0695 \\ 
 &  & AIPTW & 0.0042 & 0.0037 & 0.0047 & 0.0055 & 0.0625 \\ \hline
2 & A & KM & 0.1475 & 0.1475 & 0.1475 & 0.1475 & 0.1475 \\ 
 &  & DS & 0.0253 & 0.0258 & 0.0897 & 0.0253 & 0.0897 \\ 
 &  & IPTW KM & 0.0065 & 0.0061 & 0.0065 & 0.0740 & 0.0740 \\ 
 &  & IPTW CH & -0.0029 & -0.0034 & -0.0029 & 0.0647 & 0.0647 \\ 
 &  & Matching & 0.0080 & 0.0119 & 0.0080 & 0.0732 & 0.0732 \\ 
 &  & EL & -0.0044 & 0.0008 & -0.0044 & 0.0793 & 0.0793 \\ 
 &  & AIPTW & 0.0058 & 0.0058 & 0.0059 & 0.0074 & 0.0724 \\ \hline
2 & B & KM & 0.1449 & 0.1449 & 0.1449 & 0.1449 & 0.1449 \\ 
 &  & DS & 0.0247 & 0.0255 & 0.0891 & 0.0247 & 0.0891 \\ 
 &  & IPTW KM & 0.0091 & 0.0081 & 0.0091 & 0.0756 & 0.0756 \\ 
 &  & IPTW CH & -0.0008 & -0.0021 & -0.0008 & 0.0660 & 0.0660 \\ 
 &  & Matching & 0.0114 & 0.0101 & 0.0114 & 0.0745 & 0.0745 \\ 
 &  & EL & -0.0013 & -0.0007 & -0.0013 & 0.0833 & 0.0833 \\ 
 &  & AIPTW & 0.0086 & 0.0084 & 0.0090 & 0.0099 & 0.0753 \\ \hline
\end{tabular}
\centering
\\
\vspace{0.3cm}
\begin{minipage}{360pt}
Methods for deriving survival curves: 
KM, Standard Kaplan-Meier based approach (unadjusted); 
DS, Direct Standardization via a Cox model (G-formula); 
IPTW KM, Xie and Liu's approach;
IPTW CH, Cole and Hernan's approach; 
Matching, Propensity score matching;
EL, Empirical Likelihood approach;
AIPTW, Augmented Inverse Probability of Treatment Weighting approach.\\
\begin{tabular}{cll}
Case & Outcome model & Treatment model \\ \hline  
1 & Correct & Correct \\
2 & Included extra variables, $X_3$ and $X_6$ & 
    Included extra variables, $X_1$ and $X_4$ \\
3 & Failed to include $X_2$ & Correct \\
4 & Correct & Failed to include $X_2$ \\
5 & Failed to include $X_2$ & Failed to include $X_2$ \\ \hline 
\end{tabular}
\end{minipage}\end{table}

\begin{table}[t!]
\renewcommand\thetable{A6}
\centering
\caption{Square-root of Mean Square Error (rMSE) for AH in Group 1}
\label{tab:Group 1 AH RMSE}
\begin{tabular}{cclccccc}
\hline
&  &  & & & Case  &  & \\
Scenario & Censoring & Method & 1 & 2 & 3 & 4 & 5 \\ \hline
1 & A & KM & 0.1992 & 0.1992 & 0.1992 & 0.1992 & 0.1992 \\ 
 &  & DS & 0.1035 & 0.1067 & 0.1261 & 0.1035 & 0.1261 \\ 
 &  & IPTW KM & 0.1446 & 0.1427 & 0.1446 & 0.1507 & 0.1507 \\ 
 &  & IPTW CH & 0.1482 & 0.1461 & 0.1482 & 0.1619 & 0.1619 \\ 
 &  & Matching & 0.1707 & 0.1678 & 0.1707 & 0.1721 & 0.1721 \\ 
 &  & EL & 0.1702 & 0.1631 & 0.1702 & 0.1678 & 0.1678 \\ 
 &  & AIPTW & 0.1465 & 0.1483 & 0.1468 & 0.1351 & 0.1505 \\ \hline
1 & B & KM & 0.1993 & 0.1993 & 0.1993 & 0.1993 & 0.1993 \\ 
 &  & DS & 0.1047 & 0.1081 & 0.1279 & 0.1047 & 0.1279 \\ 
 &  & IPTW KM & 0.1482 & 0.1460 & 0.1482 & 0.1554 & 0.1554 \\ 
 &  & IPTW CH & 0.1510 & 0.1486 & 0.1510 & 0.1645 & 0.1645 \\ 
 &  & Matching & 0.1749 & 0.1720 & 0.1749 & 0.1760 & 0.1760 \\ 
 &  & EL & 0.1735 & 0.1641 & 0.1735 & 0.1706 & 0.1706 \\ 
 &  & AIPTW & 0.1494 & 0.1514 & 0.1498 & 0.1384 & 0.1546 \\ \hline
2 & A & KM & 0.2886 & 0.2886 & 0.2886 & 0.2886 & 0.2886 \\ 
 &  & DS & 0.1697 & 0.1744 & 0.2135 & 0.1697 & 0.2135 \\ 
 &  & IPTW KM & 0.2201 & 0.2152 & 0.2201 & 0.2415 & 0.2415 \\ 
 &  & IPTW CH & 0.2157 & 0.2107 & 0.2157 & 0.2330 & 0.2330 \\ 
 &  & Matching & 0.2603 & 0.2624 & 0.2603 & 0.2772 & 0.2772 \\ 
 &  & EL & 0.2687 & 0.2492 & 0.2687 & 0.2647 & 0.2647 \\ 
 &  & AIPTW & 0.2229 & 0.2261 & 0.2227 & 0.2082 & 0.2404 \\ \hline
2 & B & KM & 0.2862 & 0.2862 & 0.2862 & 0.2862 & 0.2862 \\ 
 &  & DS & 0.1688 & 0.1743 & 0.2120 & 0.1688 & 0.2120 \\ 
 &  & IPTW KM & 0.2316 & 0.2256 & 0.2316 & 0.2497 & 0.2497 \\ 
 &  & IPTW CH & 0.2266 & 0.2205 & 0.2266 & 0.2409 & 0.2409 \\ 
 &  & Matching & 0.2788 & 0.2706 & 0.2788 & 0.2837 & 0.2837 \\ 
 &  & EL & 0.2715 & 0.2551 & 0.2715 & 0.2768 & 0.2768 \\ 
 &  & AIPTW & 0.2350 & 0.2381 & 0.2347 & 0.2169 & 0.2500 \\ \hline
\end{tabular}
\centering
\\
\vspace{0.3cm}
\begin{minipage}{360pt}
Methods for deriving survival curves: 
KM, Standard Kaplan-Meier based approach (unadjusted); 
DS, Direct Standardization via a Cox model (G-formula); 
IPTW KM, Xie and Liu's approach;
IPTW CH, Cole and Hernan's approach; 
Matching, Propensity score matching;
EL, Empirical Likelihood approach;
AIPTW, Augmented Inverse Probability of Treatment Weighting approach.\\
\begin{tabular}{cll}
Case & Outcome model & Treatment model \\ \hline  
1 & Correct & Correct \\
2 & Included extra variables, $X_3$ and $X_6$ & 
    Included extra variables, $X_1$ and $X_4$ \\
3 & Failed to include $X_2$ & Correct \\
4 & Correct & Failed to include $X_2$ \\
5 & Failed to include $X_2$ & Failed to include $X_2$ \\ \hline 
\end{tabular}
\end{minipage}
\end{table}

\begin{table}[t!]
\renewcommand\thetable{A7}
\centering
\caption{Coverage Probability of 0.95 Confidence Interval for AH in Group 1}
\label{tab:Group 1 AH Coverage Probability}
\begin{tabular}{cclccccc}
\hline
&  &  & & & Case  &  & \\
Scenario & Censoring & Method & 1 & 2 & 3 & 4 & 5 \\ \hline
1 & A & KM & 0.843 & 0.843 & 0.843 & 0.843 & 0.843 \\ 
 &  & DS & 0.951 & 0.949 & 0.919 & 0.951 & 0.919 \\ 
 &  & IPTW KM & 0.935 & 0.938 & 0.935 & 0.942 & 0.942 \\ 
 &  & IPTW CH & 0.948 & 0.950 & 0.948 & 0.943 & 0.943 \\ 
 &  & Matching & 0.934 & 0.946 & 0.934 & 0.938 & 0.938 \\ 
 &  & EL & 0.936 & 0.944 & 0.936 & 0.940 & 0.940 \\ 
 &  & AIPTW & 0.947 & 0.947 & 0.949 & 0.955 & 0.959 \\ \hline
1 & B & KM & 0.843 & 0.843 & 0.843 & 0.843 & 0.843 \\ 
 &  & DS & 0.952 & 0.953 & 0.924 & 0.952 & 0.924 \\ 
 &  & IPTW KM & 0.935 & 0.941 & 0.935 & 0.943 & 0.943 \\ 
 &  & IPTW CH & 0.942 & 0.943 & 0.942 & 0.936 & 0.936 \\ 
 &  & Matching & 0.937 & 0.944 & 0.937 & 0.939 & 0.939 \\ 
 &  & EL & 0.936 & 0.946 & 0.936 & 0.942 & 0.942 \\ 
 &  & AIPTW & 0.952 & 0.953 & 0.960 & 0.964 & 0.968 \\ \hline
2 & A & KM & 0.898 & 0.898 & 0.898 & 0.898 & 0.898 \\ 
 &  & DS & 0.951 & 0.951 & 0.929 & 0.951 & 0.929 \\ 
 &  & IPTW KM & 0.948 & 0.953 & 0.948 & 0.953 & 0.953 \\ 
 &  & IPTW CH & 0.945 & 0.949 & 0.945 & 0.954 & 0.954 \\ 
 &  & Matching & 0.946 & 0.950 & 0.946 & 0.954 & 0.954 \\ 
 &  & EL & 0.945 & 0.950 & 0.945 & 0.949 & 0.949 \\ 
 &  & AIPTW & 0.954 & 0.951 & 0.956 & 0.959 & 0.966 \\ \hline
2 & B & KM & 0.901 & 0.901 & 0.901 & 0.901 & 0.901 \\ 
 &  & DS & 0.954 & 0.955 & 0.933 & 0.954 & 0.933 \\ 
 &  & IPTW KM & 0.936 & 0.943 & 0.936 & 0.951 & 0.951 \\ 
 &  & IPTW CH & 0.932 & 0.940 & 0.932 & 0.952 & 0.952 \\ 
 &  & Matching & 0.933 & 0.942 & 0.933 & 0.940 & 0.940 \\ 
 &  & EL & 0.933 & 0.940 & 0.933 & 0.944 & 0.944 \\ 
 &  & AIPTW & 0.952 & 0.952 & 0.953 & 0.955 & 0.967 \\ \hline
\end{tabular}
\centering
\\
\vspace{0.3cm}
\begin{minipage}{360pt}
Methods for deriving survival curves: 
KM, Standard Kaplan-Meier based approach (unadjusted); 
DS, Direct Standardization via a Cox model (G-formula); 
IPTW KM, Xie and Liu's approach;
IPTW CH, Cole and Hernan's approach; 
Matching, Propensity score matching;
EL, Empirical Likelihood approach;
AIPTW, Augmented Inverse Probability of Treatment Weighting approach.\\
\begin{tabular}{cll}
Case & Outcome model & Treatment model \\ \hline  
1 & Correct & Correct \\
2 & Included extra variables, $X_3$ and $X_6$ & 
    Included extra variables, $X_1$ and $X_4$ \\
3 & Failed to include $X_2$ & Correct \\
4 & Correct & Failed to include $X_2$ \\
5 & Failed to include $X_2$ & Failed to include $X_2$ \\ \hline 
\end{tabular}
\end{minipage}\end{table}

\begin{table}[t!]
\renewcommand\thetable{A8}
\centering
\caption{Median Length of 0.95 Confidence Interval for AH in Group 1}
\label{tab:Group 1 AH Median CI Length}
\begin{tabular}{cclccccc}
\hline
&  &  & & & Case  &  & \\
Scenario & Censoring & Method & 1 & 2 & 3 & 4 & 5 \\ \hline
1 & A & KM & 0.5305 & 0.5305 & 0.5305 & 0.5305 & 0.5305 \\ 
 &  & DS & 0.4022 & 0.4120 & 0.4169 & 0.4022 & 0.4169 \\ 
 &  & IPTW KM & 0.5426 & 0.5359 & 0.5426 & 0.5333 & 0.5333 \\ 
 &  & IPTW CH & 0.5651 & 0.5609 & 0.5651 & 0.5610 & 0.5610 \\ 
 &  & Matching & 0.6322 & 0.6431 & 0.6322 & 0.6211 & 0.6211 \\ 
 &  & EL & 0.6275 & 0.6058 & 0.6275 & 0.5723 & 0.5723 \\ 
 &  & AIPTW & 0.5531 & 0.5598 & 0.5617 & 0.5332 & 0.5509 \\ \hline
1 & B & KM & 0.5372 & 0.5372 & 0.5372 & 0.5372 & 0.5372 \\ 
 &  & DS & 0.4107 & 0.4208 & 0.4254 & 0.4107 & 0.4254 \\ 
 &  & IPTW KM & 0.5542 & 0.5491 & 0.5542 & 0.5456 & 0.5456 \\ 
 &  & IPTW CH & 0.5738 & 0.5690 & 0.5738 & 0.5698 & 0.5698 \\ 
 &  & Matching & 0.6451 & 0.6553 & 0.6451 & 0.6345 & 0.6345 \\ 
 &  & EL & 0.6386 & 0.6176 & 0.6386 & 0.5844 & 0.5844 \\ 
 &  & AIPTW & 0.5909 & 0.5956 & 0.6028 & 0.5708 & 0.5896 \\ \hline
2 & A & KM & 0.8093 & 0.8093 & 0.8093 & 0.8093 & 0.8093 \\ 
 &  & DS & 0.6478 & 0.6675 & 0.6835 & 0.6478 & 0.6835 \\ 
 &  & IPTW KM & 0.8620 & 0.8472 & 0.8620 & 0.8566 & 0.8566 \\ 
 &  & IPTW CH & 0.8475 & 0.8319 & 0.8475 & 0.8446 & 0.8446 \\ 
 &  & Matching & 1.0109 & 1.0265 & 1.0109 & 1.0034 & 1.0034 \\ 
 &  & EL & 0.9908 & 0.9457 & 0.9908 & 0.9177 & 0.9177 \\ 
 &  & AIPTW & 0.8611 & 0.8708 & 0.8685 & 0.8207 & 0.8620 \\ \hline
2 & B & KM & 0.8201 & 0.8201 & 0.8201 & 0.8201 & 0.8201 \\ 
 &  & DS & 0.6571 & 0.6763 & 0.6935 & 0.6571 & 0.6935 \\ 
 &  & IPTW KM & 0.8807 & 0.8674 & 0.8807 & 0.8739 & 0.8739 \\ 
 &  & IPTW CH & 0.8626 & 0.8491 & 0.8626 & 0.8605 & 0.8605 \\ 
 &  & Matching & 1.0198 & 1.0401 & 1.0198 & 1.0203 & 1.0203 \\ 
 &  & EL & 1.0118 & 0.9681 & 1.0118 & 0.9360 & 0.9360 \\ 
 &  & AIPTW & 0.9020 & 0.9101 & 0.9176 & 0.8641 & 0.9105 \\ \hline
\end{tabular}
\centering
\\
\vspace{0.3cm}
\begin{minipage}{360pt}
Methods for deriving survival curves: 
KM, Standard Kaplan-Meier based approach (unadjusted); 
DS, Direct Standardization via a Cox model (G-formula); 
IPTW KM, Xie and Liu's approach;
IPTW CH, Cole and Hernan's approach; 
Matching, Propensity score matching;
EL, Empirical Likelihood approach;
AIPTW, Augmented Inverse Probability of Treatment Weighting approach.\\
\begin{tabular}{cll}
Case & Outcome model & Treatment model \\ \hline  
1 & Correct & Correct \\
2 & Included extra variables, $X_3$ and $X_6$ & 
    Included extra variables, $X_1$ and $X_4$ \\
3 & Failed to include $X_2$ & Correct \\
4 & Correct & Failed to include $X_2$ \\
5 & Failed to include $X_2$ & Failed to include $X_2$ \\ \hline 
\end{tabular}
\end{minipage}\end{table}

\begin{table}[t!]
\renewcommand\thetable{A9}
\centering
\caption{Relative Bias for log(RAH)}
\label{tab:Difference in log(AH) Relative Bias}
\begin{tabular}{cclccccc}
\hline
&  &  & & & Case  &  & \\
Scenario & Censoring & Method & 1 & 2 & 3 & 4 & 5 \\ \hline
1 & A & KM & -0.3021 & -0.3021 & -0.3021 & -0.3021 & -0.3021 \\ 
 &  & DS & -0.0295 & -0.0308 & -0.1670 & -0.0295 & -0.1670 \\ 
 &  & IPTW KM & 0.0062 & 0.0080 & 0.0062 & -0.1404 & -0.1404 \\ 
 &  & IPTW CH & 0.0071 & 0.0090 & 0.0071 & -0.1407 & -0.1407 \\ 
 &  & Matching & 0.0087 & 0.0113 & 0.0087 & -0.1338 & -0.1338 \\ 
 &  & EL & 0.0116 & 0.0034 & 0.0116 & -0.1573 & -0.1573 \\ 
 &  & AIPTW & 0.0065 & 0.0084 & 0.0056 & 0.0022 & -0.1403 \\ \hline
1 & B & KM & -0.3047 & -0.3047 & -0.3047 & -0.3047 & -0.3047 \\ 
 &  & DS & -0.0342 & -0.0353 & -0.1710 & -0.0342 & -0.1710 \\ 
 &  & IPTW KM & 0.0034 & 0.0045 & 0.0034 & -0.1438 & -0.1438 \\ 
 &  & IPTW CH & 0.0048 & 0.0061 & 0.0048 & -0.1435 & -0.1435 \\ 
 &  & Matching & 0.0046 & 0.0065 & 0.0046 & -0.1366 & -0.1366 \\ 
 &  & EL & 0.0043 & 0.0101 & 0.0043 & -0.1605 & -0.1605 \\ 
 &  & AIPTW & 0.0039 & 0.0051 & 0.0026 & -0.0019 & -0.1443 \\ \hline
2 & A & KM & -0.2741 & -0.2741 & -0.2741 & -0.2741 & -0.2741 \\ 
 &  & DS & -0.0427 & -0.0438 & -0.1736 & -0.0427 & -0.1736 \\ 
 &  & IPTW KM & 0.0058 & 0.0058 & 0.0058 & -0.1382 & -0.1382 \\ 
 &  & IPTW CH & 0.0075 & 0.0076 & 0.0075 & -0.1362 & -0.1362 \\ 
 &  & Matching & 0.0126 & 0.0075 & 0.0126 & -0.1342 & -0.1342 \\ 
 &  & EL & 0.0329 & -0.0129 & 0.0329 & -0.1491 & -0.1491 \\ 
 &  & AIPTW & 0.0064 & 0.0074 & 0.0057 & -0.0007 & -0.1372 \\ \hline
2 & B & KM & -0.2751 & -0.2751 & -0.2751 & -0.2751 & -0.2751 \\ 
 &  & DS & -0.0477 & -0.0495 & -0.1788 & -0.0477 & -0.1788 \\ 
 &  & IPTW KM & 0.0016 & 0.0018 & 0.0016 & -0.1415 & -0.1415 \\ 
 &  & IPTW CH & 0.0036 & 0.0041 & 0.0036 & -0.1393 & -0.1393 \\ 
 &  & Matching & 0.0036 & 0.0054 & 0.0036 & -0.1328 & -0.1328 \\ 
 &  & EL & 0.0049 & 0.0074 & 0.0049 & -0.1575 & -0.1575 \\ 
 &  & AIPTW & 0.0010 & 0.0017 & -0.0003 & -0.0055 & -0.1428 \\ \hline
\end{tabular}
\centering
\\
\vspace{0.3cm}
\begin{minipage}{360pt}
Methods for deriving survival curves: 
KM, Standard Kaplan-Meier based approach (unadjusted); 
DS, Direct Standardization via a Cox model (G-formula); 
IPTW KM, Xie and Liu's approach;
IPTW CH, Cole and Hernan's approach; 
Matching, Propensity score matching;
EL, Empirical Likelihood approach;
AIPTW, Augmented Inverse Probability of Treatment Weighting approach.\\
\begin{tabular}{cll}
Case & Outcome model & Treatment model \\ \hline  
1 & Correct & Correct \\
2 & Included extra variables, $X_3$ and $X_6$ & 
    Included extra variables, $X_1$ and $X_4$ \\
3 & Failed to include $X_2$ & Correct \\
4 & Correct & Failed to include $X_2$ \\
5 & Failed to include $X_2$ & Failed to include $X_2$ \\ \hline 
\end{tabular}
\end{minipage}\end{table}

\begin{table}[t!]
\renewcommand\thetable{A10}
\centering
\caption{Square-root of Mean Square Error (rMSE) for log(RAH)}
\label{tab:Difference in log(AH) RMSE}
\begin{tabular}{cclccccc}
\hline
&  &  & & & Case  &  & \\
Scenario & Censoring & Method & 1 & 2 & 3 & 4 & 5 \\ \hline
1 & A & KM & 0.2370 & 0.2370 & 0.2370 & 0.2370 & 0.2370 \\ 
 &  & DS & 0.1175 & 0.1247 & 0.1541 & 0.1175 & 0.1541 \\ 
 &  & IPTW KM & 0.1749 & 0.1676 & 0.1749 & 0.1860 & 0.1860 \\ 
 &  & IPTW CH & 0.1750 & 0.1680 & 0.1750 & 0.1864 & 0.1864 \\ 
 &  & Matching & 0.2065 & 0.1982 & 0.2065 & 0.2078 & 0.2078 \\ 
 &  & EL & 0.3764 & 0.5153 & 0.3764 & 0.2594 & 0.2594 \\ 
 &  & AIPTW & 0.1703 & 0.1719 & 0.1708 & 0.1610 & 0.1805 \\ \hline
1 & B & KM & 0.2409 & 0.2409 & 0.2409 & 0.2409 & 0.2409 \\ 
 &  & DS & 0.1209 & 0.1286 & 0.1580 & 0.1209 & 0.1580 \\ 
 &  & IPTW KM & 0.1804 & 0.1732 & 0.1804 & 0.1919 & 0.1919 \\ 
 &  & IPTW CH & 0.1806 & 0.1735 & 0.1806 & 0.1924 & 0.1924 \\ 
 &  & Matching & 0.2124 & 0.2038 & 0.2124 & 0.2128 & 0.2128 \\ 
 &  & EL & 0.3799 & 0.4691 & 0.3799 & 0.2603 & 0.2603 \\ 
 &  & AIPTW & 0.1750 & 0.1771 & 0.1756 & 0.1654 & 0.1858 \\ \hline
2 & A & KM & 0.2636 & 0.2636 & 0.2636 & 0.2636 & 0.2636 \\ 
 &  & DS & 0.1380 & 0.1450 & 0.1862 & 0.1380 & 0.1862 \\ 
 &  & IPTW KM & 0.1888 & 0.1777 & 0.1888 & 0.2071 & 0.2071 \\ 
 &  & IPTW CH & 0.1873 & 0.1761 & 0.1873 & 0.2053 & 0.2053 \\ 
 &  & Matching & 0.2209 & 0.2158 & 0.2209 & 0.2363 & 0.2363 \\ 
 &  & EL & 0.3212 & 0.5372 & 0.3212 & 0.2611 & 0.2611 \\ 
 &  & AIPTW & 0.1808 & 0.1822 & 0.1814 & 0.1717 & 0.1991 \\ \hline
2 & B & KM & 0.2661 & 0.2661 & 0.2661 & 0.2661 & 0.2661 \\ 
 &  & DS & 0.1423 & 0.1508 & 0.1920 & 0.1423 & 0.1920 \\ 
 &  & IPTW KM & 0.1996 & 0.1891 & 0.1996 & 0.2170 & 0.2170 \\ 
 &  & IPTW CH & 0.1977 & 0.1871 & 0.1977 & 0.2150 & 0.2150 \\ 
 &  & Matching & 0.2378 & 0.2249 & 0.2378 & 0.2406 & 0.2406 \\ 
 &  & EL & 0.4113 & 0.4966 & 0.4113 & 0.2877 & 0.2877 \\ 
 &  & AIPTW & 0.1935 & 0.1960 & 0.1939 & 0.1831 & 0.2110 \\ \hline
\end{tabular}
\centering
\\
\vspace{0.3cm}
\begin{minipage}{360pt}
Methods for deriving survival curves: 
KM, Standard Kaplan-Meier based approach (unadjusted); 
DS, Direct Standardization via a Cox model (G-formula); 
IPTW KM, Xie and Liu's approach;
IPTW CH, Cole and Hernan's approach; 
Matching, Propensity score matching;
EL, Empirical Likelihood approach;
AIPTW, Augmented Inverse Probability of Treatment Weighting approach.\\
\begin{tabular}{cll}
Case & Outcome model & Treatment model \\ \hline  
1 & Correct & Correct \\
2 & Included extra variables, $X_3$ and $X_6$ & 
    Included extra variables, $X_1$ and $X_4$ \\
3 & Failed to include $X_2$ & Correct \\
4 & Correct & Failed to include $X_2$ \\
5 & Failed to include $X_2$ & Failed to include $X_2$ \\ \hline 
\end{tabular}
\end{minipage}\end{table}

\begin{table}[t!]
\renewcommand\thetable{A11}
\centering
\caption{Coverage Probability of 0.95 Confidence Interval for RAH}
\label{tab:Difference in log(AH) Coverage Probability}
\begin{tabular}{cclccccc}
\hline
&  &  & & & Case  &  & \\
Scenario & Censoring & Method & 1 & 2 & 3 & 4 & 5 \\ \hline
1 & A & KM & 0.767 & 0.767 & 0.767 & 0.767 & 0.767 \\ 
 &  & DS & 0.941 & 0.944 & 0.844 & 0.941 & 0.844 \\ 
 &  & IPTW KM & 0.945 & 0.952 & 0.945 & 0.915 & 0.915 \\ 
 &  & IPTW CH & 0.950 & 0.953 & 0.950 & 0.917 & 0.917 \\ 
 &  & Matching & 0.941 & 0.952 & 0.941 & 0.925 & 0.925 \\ 
 &  & EL & 0.977 & 0.985 & 0.977 & 0.931 & 0.931 \\ 
 &  & AIPTW & 0.969 & 0.971 & 0.971 & 0.971 & 0.938 \\ \hline
1 & B & KM & 0.762 & 0.762 & 0.762 & 0.762 & 0.762 \\ 
 &  & DS & 0.946 & 0.950 & 0.853 & 0.946 & 0.853 \\ 
 &  & IPTW KM & 0.945 & 0.949 & 0.945 & 0.908 & 0.908 \\ 
 &  & IPTW CH & 0.950 & 0.951 & 0.950 & 0.910 & 0.910 \\ 
 &  & Matching & 0.947 & 0.953 & 0.947 & 0.918 & 0.918 \\ 
 &  & EL & 0.974 & 0.986 & 0.974 & 0.931 & 0.931 \\ 
 &  & AIPTW & 0.977 & 0.977 & 0.980 & 0.983 & 0.956 \\ \hline
2 & A & KM & 0.753 & 0.753 & 0.753 & 0.753 & 0.753 \\ 
 &  & DS & 0.939 & 0.943 & 0.829 & 0.939 & 0.829 \\ 
 &  & IPTW KM & 0.951 & 0.962 & 0.951 & 0.914 & 0.914 \\ 
 &  & IPTW CH & 0.951 & 0.963 & 0.951 & 0.914 & 0.914 \\ 
 &  & Matching & 0.954 & 0.960 & 0.954 & 0.916 & 0.916 \\ 
 &  & EL & 0.976 & 0.980 & 0.976 & 0.930 & 0.930 \\ 
 &  & AIPTW & 0.967 & 0.969 & 0.968 & 0.969 & 0.932 \\ \hline
2 & B & KM & 0.761 & 0.761 & 0.761 & 0.761 & 0.761 \\ 
 &  & DS & 0.939 & 0.942 & 0.829 & 0.939 & 0.829 \\ 
 &  & IPTW KM & 0.940 & 0.945 & 0.940 & 0.902 & 0.902 \\ 
 &  & IPTW CH & 0.940 & 0.944 & 0.940 & 0.904 & 0.904 \\ 
 &  & Matching & 0.942 & 0.943 & 0.942 & 0.922 & 0.922 \\ 
 &  & EL & 0.973 & 0.984 & 0.973 & 0.925 & 0.925 \\ 
 &  & AIPTW & 0.967 & 0.967 & 0.968 & 0.967 & 0.937 \\ \hline
\end{tabular}
\centering
\\
\vspace{0.3cm}
\begin{minipage}{360pt}
Methods for deriving survival curves: 
KM, Standard Kaplan-Meier based approach (unadjusted); 
DS, Direct Standardization via a Cox model (G-formula); 
IPTW KM, Xie and Liu's approach;
IPTW CH, Cole and Hernan's approach; 
Matching, Propensity score matching;
EL, Empirical Likelihood approach;
AIPTW, Augmented Inverse Probability of Treatment Weighting approach.\\
\begin{tabular}{cll}
Case & Outcome model & Treatment model \\ \hline  
1 & Correct & Correct \\
2 & Included extra variables, $X_3$ and $X_6$ & 
    Included extra variables, $X_1$ and $X_4$ \\
3 & Failed to include $X_2$ & Correct \\
4 & Correct & Failed to include $X_2$ \\
5 & Failed to include $X_2$ & Failed to include $X_2$ \\ \hline 
\end{tabular}
\end{minipage}\end{table}

\begin{table}[t!]
\renewcommand\thetable{A12}
\centering
\caption{Median Length of 0.95 Confidence Interval for log(RAH)}
\label{tab:Difference in log(AH) Median CI Length}
\begin{tabular}{cclccccc}
\hline
&  &  & & & Case  &  & \\
Scenario & Censoring & Method & 1 & 2 & 3 & 4 & 5 \\ \hline
1 & A & KM & 0.5885 & 0.5885 & 0.5885 & 0.5885 & 0.5885 \\ 
 &  & DS & 0.4550 & 0.4808 & 0.4500 & 0.4550 & 0.4500 \\ 
 &  & IPTW KM & 0.6735 & 0.6486 & 0.6735 & 0.6386 & 0.6386 \\ 
 &  & IPTW CH & 0.6834 & 0.6583 & 0.6834 & 0.6515 & 0.6515 \\ 
 &  & Matching & 0.7857 & 0.7829 & 0.7857 & 0.7417 & 0.7417 \\ 
 &  & EL & 1.2383 & 1.7801 & 1.2383 & 0.7029 & 0.7029 \\ 
 &  & AIPTW & 0.6988 & 0.7077 & 0.7042 & 0.6767 & 0.6662 \\ \hline
1 & B & KM & 0.5974 & 0.5974 & 0.5974 & 0.5974 & 0.5974 \\ 
 &  & DS & 0.4642 & 0.4909 & 0.4585 & 0.4642 & 0.4585 \\ 
 &  & IPTW KM & 0.6850 & 0.6622 & 0.6850 & 0.6512 & 0.6512 \\ 
 &  & IPTW CH & 0.6951 & 0.6698 & 0.6951 & 0.6609 & 0.6609 \\ 
 &  & Matching & 0.7985 & 0.7990 & 0.7985 & 0.7545 & 0.7545 \\ 
 &  & EL & 1.2398 & 1.8078 & 1.2398 & 0.7166 & 0.7166 \\ 
 &  & AIPTW & 0.7670 & 0.7727 & 0.7771 & 0.7442 & 0.7313 \\ \hline
2 & A & KM & 0.6516 & 0.6516 & 0.6516 & 0.6516 & 0.6516 \\ 
 &  & DS & 0.5291 & 0.5561 & 0.5279 & 0.5291 & 0.5279 \\ 
 &  & IPTW KM & 0.7460 & 0.7114 & 0.7460 & 0.7142 & 0.7142 \\ 
 &  & IPTW CH & 0.7398 & 0.7057 & 0.7398 & 0.7095 & 0.7095 \\ 
 &  & Matching & 0.8691 & 0.8676 & 0.8691 & 0.8278 & 0.8278 \\ 
 &  & EL & 1.3189 & 1.9346 & 1.3189 & 0.7882 & 0.7882 \\ 
 &  & AIPTW & 0.7489 & 0.7533 & 0.7536 & 0.7232 & 0.7175 \\ \hline
2 & B & KM & 0.6594 & 0.6594 & 0.6594 & 0.6594 & 0.6594 \\ 
 &  & DS & 0.5342 & 0.5652 & 0.5335 & 0.5342 & 0.5335 \\ 
 &  & IPTW KM & 0.7580 & 0.7227 & 0.7580 & 0.7250 & 0.7250 \\ 
 &  & IPTW CH & 0.7514 & 0.7153 & 0.7154 & 0.7200 & 0.7200 \\ 
 &  & Matching & 0.8838 & 0.8794 & 0.8838 & 0.8394 & 0.8394 \\ 
 &  & EL & 1.3621 & 1.9353 & 1.3621 & 0.7918 & 0.7918 \\ 
 &  & AIPTW & 0.7989 & 0.8026 & 0.8049 & 0.7745 & 0.7659 \\ \hline
\end{tabular}
\centering
\\
\vspace{0.3cm}
\begin{minipage}{360pt}
Methods for deriving survival curves: 
KM, Standard Kaplan-Meier based approach (unadjusted); 
DS, Direct Standardization via a Cox model (G-formula); 
IPTW KM, Xie and Liu's approach;
IPTW CH, Cole and Hernan's approach; 
Matching, Propensity score matching;
EL, Empirical Likelihood approach;
AIPTW, Augmented Inverse Probability of Treatment Weighting approach.\\
\begin{tabular}{cll}
Case & Outcome model & Treatment model \\ \hline  
1 & Correct & Correct \\
2 & Included extra variables, $X_3$ and $X_6$ & 
    Included extra variables, $X_1$ and $X_4$ \\
3 & Failed to include $X_2$ & Correct \\
4 & Correct & Failed to include $X_2$ \\
5 & Failed to include $X_2$ & Failed to include $X_2$ \\ \hline 
\end{tabular}
\end{minipage}\end{table}

\end{document}